\documentclass[prd,onecolumn,notitlepage,preprintnumbers,nofootinbib,prl,superscriptaddress,10pt]{revtex4-1}

\def\l@subsubsection#1#2{}
\def\l@subsubsubsection#1#2{}
\makeatother

\usepackage[utf8]{inputenc}
\usepackage{slashed}
\usepackage{enumitem}
\usepackage{graphicx,amssymb,amsmath,amsthm,amsfonts,epsfig,epsf}
\usepackage[usenames,dvipsnames,svgnames,table]{xcolor}
\usepackage{epstopdf}
\definecolor{darkred}{rgb}{0.5,0,0}
\usepackage{dcolumn}
\usepackage{latexsym}
\usepackage{rotating}
\usepackage{longtable}

\usepackage{enumerate}
\usepackage{tensor,multirow}
\usepackage{url}
\usepackage[linktocpage]{hyperref}

\def\be{\begin{equation}}
\def\ee{\end{equation}}
\newcommand{\bea}{\begin{eqnarray}}
\newcommand{\eea}{\end{eqnarray}}

\def\ba{\begin{align}}
\def\ea{\end{align}}

\DeclareMathOperator\diag{diag}

\allowdisplaybreaks

\begin{document}

\title{{\bf Chiral anomalies induced by gravitational waves}}

\author{Adrian del Rio}
\affiliation{Institute for Gravitation and the Cosmos, Physics Department,
Penn State, University Park, PA 16802-6300, USA.}

\begin{abstract}
Chiral symmetries in field theory are typically affected by an anomaly in the quantum theory. This anomaly emerges when one introduces an interaction with a Yang-Mills or gravitational background.  Physical applications of this quantum effect have  been traditionally connected to topological questions of the background field and the study of instantons. We show here how one can alternatively  find situations of physical interest that only involve ordinary, but dynamical solutions of the background field equations. More precisely, we show that solutions to the Einstein (Maxwell) equations are able to trigger the chiral anomaly if and only if they admit a flux of gravitational (electromagnetic) radiation with net circular polarization. 
As a consequence, astrophysical systems that admit such radiation  spontaneously generate a  flux of  particles with net helicity from the quantum vacuum. 
\end{abstract}

\date{\today}

\maketitle

\section{Introduction}

It is well-known that  strong gravitational fields can affect the vacuum fluctuations of  quantum fields and lead to important physical phenomena. From the pioneer work of Parker on particle creation in expanding universes \cite{Parker68}, to the subsequent discover by Hawking of thermal emission during a gravitational collapse \cite{Hawking75}, different phenomena of quantum origin  can arise if a quantum field propagates on a dynamical, gravitational background. One of these quantum effects is related to the emergence of  anomalies due to spacetime curvature. 

An anomaly  is   understood as  the failure of some Noether symmetry of a classical field theory to persist after the quantization. More precisely,  when the classical conservation law of a  Noether current breaks down in the quantum theory, the associated symmetry is said to be anomalous \cite{Bertlmann:1996xk}. Their discovery was initiated in the late 60's with the fermion axial or chiral anomaly \cite{Adler1969, BellJackiw1969},  motivated with the aim of  understanding the observed phenomenon of the pion decay into two photons.  Since then, the field has grown enormously, leading to the discovery of many more anomalous symmetries of diverse nature (conformal, gauge, etc), and to a rich interplay with differential geometry and topology \cite{Nakahara}.  From a physical viewpoint, anomalies  were found useful  to address key conceptual questions in the standard model of particles (U(1) problem, 
 strong CP violation in QCD \cite{Schwartz2014}) and cosmology (baryogenesis). Our goal in this paper is to point out and develop an unexplored aspect of  chiral anomalies that, remarkably, turns out to have a  simple physical interpretation, and could lead to new physical applications in gravity and electrodynamics. 

For definiteness, let $\psi(x)$ be a Dirac field interacting with a classical Yang-Mills background of field strength $F_{ab}$ in Minkowski spacetime $(\mathbb R^4, \eta_{ab})$,  with coupling constant $g$. Let $\gamma^a$ be the Dirac matrices, and let $\gamma^5:=i \gamma^0 \gamma^1\gamma^2\gamma^3$ be the chiral matrix. In the massless limit, the standard action of this theory possesses a (global)  Noether symmetry generated by the transformation $\psi(x) \to e^{i\gamma^5 \theta}\psi(x)$, $\theta\in \mathbb R$, that leads to a Noether current:
$
j^a_5(x)=\bar \psi(x) \gamma^a \gamma^5 \psi(x)\, .
$
This is the well-known (abelian) chiral symmetry.
This current is conserved for solutions $\psi(x)$ of the Dirac equation of motion, $\nabla_a j^a_5 (x) \approx 0$.  In the quantum theory, however, off-shell contributions yield 
\bea
\left< \nabla_a j^a_5\right> = \frac{-g^2}{16\pi^2} {\rm Tr}\, F_{ab}{^*F}^{ab}\neq 0\, , \label{fermionanomaly}
\eea
thus spoiling the classical conservation law. This is the indication that the classical symmetry is anomalous in the quantum theory. 
Denoting  by $\psi_L=1/2(\mathbb I+\gamma^5)\Psi$ and $\psi_R=1/2(\mathbb I-\gamma^5)\Psi$ the right-handed and left-handed chiral sectors of the Dirac field, respectively, the associated Noether charge can be written as
$
Q_5(t)=\int_{\Sigma} d^3x \sqrt{h} (\psi_R^{\dagger} \psi_R - \psi_L^{\dagger} \psi_L)\, ,
$
which is a measure of the net difference between right-handed  (positive-helicity  particles plus negative-helicity antiparticles) and left-handed fermions  (hegative-helicity particles plus positive-helicity antiparticles). While this difference is  preserved by the  equation of motion for classical fields, $\dot Q_5(t)\approx 0$, the emergence of the anomaly indicates that quantum fluctuations are able to induce a change in time given by
\bea
\left<\hat Q_5(t_2)\right> - \left<\hat Q_5(t_1)\right> = \int_{[t_1,t_2]\times \Sigma} d^4x \sqrt{-\eta} \left< \nabla_a j^a_5\right>=\frac{-g^2}{16\pi^2} \int_{[t_1,t_2]\times \Sigma} d^4x \sqrt{-\eta} {\rm Tr}\, F_{ab}{^*F}^{ab} \, ,
\eea
 if and only if  the RHS of this equation is non-vanishing. 
 What does the anomalous time dependence of the chiral charge imply physically?  Using S-matrix theory and Bogoliubov transformations,  it can be explicitly shown   \cite{Christ1980} that the Yang-Mills field creates and destroys fermions in such a way that $\left<Q_5(t_2)\right> - \left<Q_5(t_1)\right>$, which is  a measure of asymmetric particle creation, is given precisely by the amount predicted  by the RHS of the previous equation. Thus,  the anomalous temporal evolution of the chiral charge is physically interpreted as a phenomenon of asymmetric particle creation by a dynamical background: a non-trivial gauge field  is able to excite  spontaneously  a net number of right-handed fermions over left-handed ones from the quantum vacuum, or viceversa. 
 
In a similar fashion, it has been recently shown \cite{AdRNS2017a,AdRNS2017b,AdRNS2018a, AdRNS2018b} that the analogous chiral symmetry in  electrodynamics, most popularly known as electric-magnetic duality symmetry of source-free Maxwell equations, suffers from a similar anomaly when a non-trivial spacetime background $(\mathbb R^4, g_{ab})$ is introduced. More precisely, under a chiral rotation ${^\pm F}(x) \to e^{\mp i\theta}\, {^\pm F}(x)$ of the self-dual and anti self-dual sectors of the electromagnetic field, ${^\pm F}_{ab}=\frac{1}{2}[F_{ab}\pm i{^*}F_{ab}]$, the usual action for the source-free Maxwell theory  remains invariant. This leads to a Noether current that, for solutions of the field equations, reads $J_5^a\approx A_b {^*F}^{ab}-Z_b F^{ab}$. Although classically conserved, it was found in \cite{AdRNS2017a,AdRNS2017b,AdRNS2018a, AdRNS2018b} that quantum fluctuations of the electromagnetic field produce
\bea
\left< \nabla_a J^a_5\right> = \frac{-1}{96\pi^2} R_{abcd}{^*R}^{abcd} \, , \label{EBanomaly}
\eea
if spacetime curvature ${R_{abc}}^d$ is considered. The associated Noether charge can be expressed as the difference between right-handed and left-handed  photons, and, while   Maxwell equations guarantee that $\dot Q_5\approx 0$ for the classical function $Q_5(t)$, off-shell contributions can spontaneously make this quantity change in time if and only if the RHS of
\bea 
\left<\hat Q_5(t_2)\right> - \left<\hat Q_5(t_1)\right> =\frac{-1}{96\pi^2} \int_{[t_1,t_2]\times \Sigma} d^4x \sqrt{-g} R_{abcd}{^*R}^{abcd}, \label{gravcp}
\eea
 is different from zero.  Again, the physical picture is analogous to the fermion case: a non-trivial gravitational background would be responsible to create spontaneosly a difference in the number of right- and left-handed circularly polarized photons from the quantum vacuum. 

In general, the physical interpretation of chiral anomalies is strongly associated to the phenomenon of ``level-crossing''. The Hamiltonian of the quantum field, which determines the energy of field modes, depends on the background field. Then, a non-trivial temporal evolution of the latter can make a positive-chirality mode with initial negative energy transform into a positive-chirality mode with final positive energy \cite{Kiskis1978, Christ1980}. In other words, the dynamical evolution of the Yang-Mills or gravitational field can reverse the helicity of field modes, producing as a result a net creation of helicity from the quantum vacuum: more particles of one helicity are created than particles of the opposite helicity. 

The question we want to answer here is: what are the physical spacetime backgrounds that can induce this level-crossing in the helicity of field modes? How is this dynamical evolution supposed to be? In the Yang-Mills case an important historical role has  been played by instantons   \cite{Jackiw77, Shifman1994}. Instantons are classical solutions to the Euclidean field equations of a non-abelian gauge theory which exhibit a non-trivial topology in the manifold of field configurations. Physically, they are  interpreted as amplitudes that quantify quantum-mechanical transitions between topologically-inequivalent vacua in the Hilbert space of  gauge fields \cite{JackiwRebbi77, CallanDashenGross1976, Jackiw77, BitarChang1978}. Their use played a fundamental role in the 70-80 in addressing several problems of the standard model of particles and QCD, most notably the $U(1)$ problem \cite{tHooft76}.  
But what about the gravitational case? Analogous solutions of the Euclidean Einstein's equations are also known for a long time \cite{Hawking1977, EguchiHanson1979} and were baptised as gravitational instantons. While mathematically these solutions have a rich structure \cite{Dunajski2010},  their physical interpretation could be regarded as exotic, since it relies on quantum-gravity issues \cite{THOOFT1989517}. If one is only interested in studying applications of chiral anomalies, it can be more natural   to look instead for more realistic spacetimes, i.e. for
 {\it ordinary}, Lorentzian solutions of Einsteins equations directly. 
  
To our knowledge, this problem has not been proposed in the literature before. Since it is known that the integrand in (\ref{gravcp}) is locally a total derivative, a priori one  expects that only non-trivial spacetime topologies can produce non-vanishing contributions,  thereby the historical interest in instantons. However, in gravity this is not quite true because, at least for  asymptotically flat spacetimes, the  boundary of the spacetime, called null infinity \cite{Penrose1963}, is non-trivial and can provide a contribution  by means of the flux of gravitational  waves (GWs) that enters/exits the spacetime. The goal of this paper is to fill this gap, and to open a new window for applications of chiral  anomalies that go beyond the realm of topology or instanton calculus. 

Our main result will be to show that chiral anomalies are intimately  related to the  circular polarization state of ordinary  gravitational radiation in the spacetime background. More precisely, we shall prove that:
\bea
\left<\hat Q_5(scri+)\right> - \left<\hat Q_5(scri-)\right>  = \int_0^{\infty} \frac{d\omega \omega^3}{24\pi^3} \sum_{\ell m} \left[| h_+^{\ell m}(\omega)-ih^{\ell m}_{\times}(\omega)|^2-|h_+^{\ell m}(\omega) + ih^{\ell m}_{\times}(\omega)|^2 \right] \, ,
\eea
where $h_+$, $h_{\times}$ denote the two GW linear polarization modes that reach future null infinity, emitted by an isolated gravitational source that is stationary at both past and future timelike infinities, but otherwise arbitrary.  What this formula is indicating is that the more right(left)-handed gravitational radiation is emitted by a system, the more right(left)-handed particles will be excited from the quantum vacuum through the mechanism that produces the chiral anomaly. This is  a  realistic gravitational setting,  with a clear and unambiguous physical meaning. 

That an asymptotically flat spacetime background must emit gravitational waves in order to induce the quantum anomaly is an indication that only {\it dynamical solutions} of Einstein's equations are relevant in this question.
\footnote{But not all dynamical spacetimes produce a non-vanishing Chern-Pontryagin. One can easily check that Fridmann-Lemaitre-Robertson-Walker metrics produce a zero result.} 
This should come with no surprise in light of the previous physical interpretation of anomalies in terms of asymmetric particle creation, since only {\it dynamical gravitational fields} are able to spontaneously create particles (and hence helicity) from the quantum vacuum. On the other hand, because the study dynamical solutions of Einstein's equations is a rather involved issue that typically requires numerical techniques, this  could  explain why only instantonic solutions (which are known in closed form) have only been considered so far in the study of chiral anomalies. 

In a given sense the result that we  obtain shares some parallelisms and  interpretations with instantons. Namely, our result can be understood as tunneling between  degenerate vacua of the  asymptotically flat spacetime (associated to the degeneracy of gravitational connections at future null infinity \cite{Ashtekar1981, Ashtekar1981b}), but in this case these transitions are produced {\it classically} simply by a flux of  gravitational waves crossing null infinity (i.e. not through the usual quantum-mechanical tunneling). However, important conceptual differences exist. For instance,  our result indicates that any spacetime  that its manifold is homeomorphic to  $\mathbb R^4$, but such that its metric is  deformed with respect to Minkowski so as to allow the presence of circularly polarized gravitational radiation (i.e. to allow for curvature, in a specific form) will be able to induce the quantum anomaly and hence level-crossing of modes. Thus, our result has nothing to do with topological or global questions, but rather to the geometry of the spacetime\footnote{In the context of the Atiyah-Patodi-Singer index theorem, this geometric contribution is essentially the one that must be subtracted from the Chern-Pontryagin integral in order to recover a topological quantity for  manifolds with boundary, see \cite{EHG}}. 
Moreover, it seems unlikely that these results could be simply obtained by a wick rotation of an instanton solution, since the former are purely dynamical (gravitational waves), and hence it does not look that they could be recovered from ``static'' euclidean solutions  by any analytical continuation. 

 In light of the systematic detections of gravitational waves by the interferometers LIGO-Virgo in the last years \cite{LIGOVirgoSummary}, and given that gravitational backgrounds emitting these waves are  intimately related to chiral anomalies, it is  important to discuss astrophysical settings where these quantum effects could play a role in the underlying physics. What is the possible phenomenology that one could predict? This will be studied in detail in a separate paper. The present paper is an extended and a detailed discussion of the theoretical results presented already in \cite{dRSGMAFNS}.

We shall work with 4-dimensional spacetimes and  the Levi-Civita connection. We  follow Wald's  \cite{Wald84} sign conventions. Namely, the metric signature is $(-,+,+,+)$, the Riemann tensor is defined by $[\nabla_a,\nabla_b]v_c=:R_{abc}^{\hspace{0.45cm}d}v_d$ for any covector field $v_d$, the Ricci tensor is  $R_{ab}:=R^{c}_{\hspace{0.15cm}acb}$, and the scalar curvature is $R:=g^{ab}R_{ab}$. Unless otherwise stated, all tensor fields will be considered smooth. In section II we use units in which $\hbar=c=1$, while in section III we use units in which $G = c = 1$.

\section{A simple case: the electromagnetic analogue}

Although we are ultimately interested in understanding the chiral anomaly induced by a gravitational background, the usual technical complications associated with the non-linearities of the gravitational field and  Einstein's equations makes necessary   working first with a simpler model. Consequently, let us focus in this section on the original Adler-Bell-Jackiw chiral anomaly  \cite{Adler1969, BellJackiw1969}, which is the electromagnetic analogue. The above complications  are  avoided due to the simplicity of Maxwell theory. More importantly, the final result  will have such a simple physical interpretation that will guide us in the gravitational case.

\subsection{Setup and main calculation} \label{IIA}

The Adler-Bell-Jackiw chiral anomaly is the anomalous non-conservation of the Noether current   associated to the chiral symmetry of a Dirac spinor $\Psi(x)$. This anomaly arises when the spinor  interacts with a background electromagnetic field $F_{ab}(x)$. The explicit expression for this anomaly can be read off from  (\ref{fermionanomaly}) if we identify the gauge group with $U(1)$ and $g^2\to e^2$. 
As emphasized in the introduction, the   quantity of major physical interest  is    the Noether charge, whose failure to be preserved in time is determined by the integral over all spacetime $M$:
\bea
\left<\hat Q_5(scri+)\right> - \left<\hat Q_5(scri-)\right>  =-\frac{e^2}{8\pi^2}\int_{M}  F\wedge F  = -\frac{e^2}{16\pi^2}\int_{M} d^4x \sqrt{-\eta} F_{ab} {^*F}^{ab} \label{intF}\, .
\eea
Our goal is to determine which class of electromagnetic backgrounds (i.e. solutions $F_{ab}$ of Maxwell equations) produce a non-vanishing chiral anomaly. To achieve this we need to analyze this integral and to study under which conditions it is not zero. 
We  assume that this electromagnetic background $F_{ab}$ is produced by   some electromagnetic sources $J^a$, that are smooth and with spatial compact support, but otherwise arbitrary. 

First of all, it is not difficult to prove that for stationary solutions of Maxwell equations, $dF=0$, $d{^*F}={^*J}$ , eq. (\ref{intF}) above is identically zero. To see this, let us take a cartesian coordinate system $\{t, \vec x\}$.  A stationary Maxwell field is a solution of Maxwell's equations that remains invariant under time translations, i.e. a solution that satisfies  $\mathcal L_k F_{ab}=0$, where $\mathcal L_k$ denotes the Lie derivative along the generator of   infinitesimal time-translations, $k=\partial/\partial t$. This condition is equivalent to $k^a F_{ab}=\nabla_b \Lambda$, where $\Lambda$ is a function, traditionally called the electrostatic potential.
The integral of interest can be now rewritten as
\bea
\int_{M} d^4x \sqrt{-\eta} F_{ab} {^*F}^{ab} &  = & - 4\int_{-\infty}^{\infty}dt \int_{\mathbb R^3} d^3 \vec x \, k^a F_{ab} k_c{^*F}^{cb}= -4\int_{-\infty}^{\infty}dt \int_{\mathbb R^3} d^3 \vec x \, \nabla_b \Lambda k_c{^*F}^{cb} \nonumber\\
&  = &  - 4\int_{-\infty}^{\infty}dt \int_{\mathbb R^3} d^3 \vec x \,  \nabla_b(\Lambda\,  k_c {^*F}^{cb}) \, ,
\eea 
where in the first equality we used the 3+1 decomposition of the metric, $\eta^{ab}=-k^a k^b+h^{ab}$, and $^*F_{ab}=\frac{1}{2}\epsilon_{abcd}F^{cd}$ to write $h^{ab}h^{cd}F_{ac}{^*F_{bd}}= - 2 k^a F_{ab} k_c{^*F}^{cb}$; while in the third equality  we used  Maxwell equation $\nabla_a {^*F}^{ab}=0$. Assuming standard  fall-off conditions for the magnetic field and electrostatic potential at spatial infinity, $k^a{^*F}_{ab}\nabla^b r \sim 1/r^3$,  $\Lambda\sim 1/r$,   the final result is zero.

We must look then for non-stationary solutions to Maxwell's equations.  To guarantee convergence for the integral in time in eq. ((\ref{intF})), we shall assume that both at early and late times the solution of Maxwell equations approaches a stationary configuration. 
\footnote{At early times the electromagnetic field is stationary for all $\vec x\in \mathbb R^3$ and using the same arguments as above we get   $ \int_{\mathbb R^3} d^3 \vec x \,  \nabla_b(\Lambda\,  k_c {^*F}^{cb})=0$. At late times $t$  the field is stationary only in a spacelike open region $U(t)\subset \mathbb R^3$ of radius $r(t)$ that does not intersect the electromagnetic waves generated during the intermediate non-stationary period. Because the waves propagate to future infinity, $r(t)=t+const$, then $U(t\to \infty)\to \mathbb R^3$ and we  find  $ \int_{U(t)} d^3 \vec x \,  \nabla_b(\Lambda\,  k_c {^*F}^{cb})=   \int d\mathbb S^2  \, r(t)^2 \Lambda(r(t))\,  k_c {^*F}^{cb}(r(t))\nabla_b r\sim t^{-2}$ as $t\to \infty$, which guarantees convergence of the integral.}
During the non stationarity period, the dynamics of the electromagnetic sources will  generate outgoing radiation propagating to infinity (we will assume no incoming electromagnetic radiation for simplicity). The study of outgoing radiation is most conveniently carried out  within the framework of asymptotically Minwkoski spacetimes  \cite{Geroch77, Ashtekar14, Ashtekar1987}. A detailed summary of this topic can be found in Appendix B, and we will provide the key points here.
Let (${\mathbb R}^4$, $ \hat\eta_{ab}$) denote our physical, Minkowski spacetime, and let ($M$, $  \eta_{ab}$) be  the unphysical spacetime constructed from the physical one by a standard conformal compactification.\footnote{Because we shall be working  with the unphysical spacetime all the time, we use the hat symbol to denote quantities associated to the physical spacetime in order to avoid its use later.} 
The unphysical metric is related to the physical one by an ordinary conformal transformation: $ \eta_{ab}=\Omega^2(x)  \hat\eta_{ab}$. On the other hand, the unphysical manifold is just the physical one together with  additional points attached smoothly to it: $M= {\mathbb R}^4\cup \mathcal J$. The set of all these new points constitute a null hypersurface $\mathcal J$, locally characterized by the condition $\Omega=0$, and with null normal $\eta^{ab}\nabla_b \Omega$. Physically, they  represent the ``points of (null) infinity'', i.e. the points that can be asymptotically reached in the original spacetime by following outgoing, null geodesics. 

The importance of  this construction is that it  allows to apply ordinary techniques in differential geometry to study the behaviour of  fields in a neigbourhood of infinity (which now is just a boundary of the  spacetime manifold). To do the calculation of interest, equation (\ref{intF}), one further needs  to carry the tensors of the original spacetime to the unphysical one. This is straightforward due to he invariance of the electromagnetic field under conformal transformations, $\hat F_{ab}=F_{ab}$, $\hat A_a=A_a$. Thus,
\bea
 -\frac{1}{2}\int_{{\mathbb R}^4} d^4x \sqrt{-\hat \eta} \hat F_{ab} {^*\hat F}^{ab}=\int_{{\mathbb R}^4}  \hat F\wedge \hat F = \int_{M}   F\wedge  F = -\frac{1}{2}\int_{M} d^4x \sqrt{- \eta}  F_{ab} {^* F}^{ab} \, .
\eea

The key point  now is to notice that, mathematically, $p_1(F)=-\frac{1}{8\pi^2}F\wedge F$ is an invariant polynomial \cite{Nakahara}.
The Chern-Weil Theorem from the theory of characteristic classes  (see Theorem 11.1 in \cite{Nakahara}, for instance)  states  that the difference between two invariant polynomials, $p_1( F)-p_1( F')$, associated to  two  connection 1-forms, $A$ and $A'$, is exact and determined by the transgression term $Q(A,A')$  \cite{Nakahara,EHG}:
\bea
p_1(F)-p_1(F')=dQ(A,A') \, .\label{ChernWeil}
\eea
Because the spacetime $(\mathbb R^4,\hat \eta_{ab})$ is  trivial from the topological viewpoint, it admits a global flat connection $\hat A'$. Due to conformal invariance and continuity, we have an electromagnetic potential that is pure gauge globally:  $A'=d\alpha$.
Then,  the difference $\theta=A-A'$ represents the same physical electromagnetic potential\footnote{The advatadge of working with $\theta$ rather than  with $A$ directly is that the former is manifestly gauge invariant, while $A$ is not.  It is customary in the literature to set $A'=0$, but this can be  misleading during the calculation given that intermediate formulas would not have a manifestly gauge-invariant form.}. The transgression term can be evaluated following its definition (see \cite{Nakahara,EHG} for details) and reads:
\bea
Q(A,A')
= -\frac{1}{8\pi^2}  \theta \wedge F \, . \label{transgression}
\eea 
It is straightforward now to check that $p_1(F')=p_1(0)=0$. Then, by integrating (\ref{ChernWeil}) and applying Stokes Theorem we get 
\bea
\int_{M} p_1(F)=\int_{\mathcal J} i^*Q(A,A')= -\frac{1}{16\pi^2}\int_{\mathcal J} \theta_a  F_{bc}\epsilon^{abc} du\, d\mathbb S^2  \label{preliminaryresult} \, ,
\eea
where the boundary of the unphysical manifold is $\partial M=\mathcal J$;  $i^*$ denotes the pullback of the  inclusion map $i: \mathcal J \hookrightarrow M$; and $du\, d\mathbb S^2$ is the canonical integration measure on $\mathcal J\approx \mathbb R\times \mathbb S^2$.
Note that the RHS is manifestly gauge-invariant, as it must be in view of the LHS.

This result can be further simplified if we work in a Newman-Penrose  basis \cite{NP62}. The electromagnetic field has 6 physical degrees of freedom per spacetime point that can be described with  3 complex scalars (these are analogous to the Weyl scalars in the gravitational case \cite{Chandrasekhar85}). These scalars are the components of the tensor  $F_{ab}$ in a  null tetrad $\{\ell^a, n^a, m^a, \bar m^a\}$.  Without loss of generality, we can take $n^a$ such that it equals $\eta^{ab}\nabla_b\Omega$ at $\Omega=0$, i.e. such that it is normal  to the hypersurface $\mathcal J$. Then $\ell^a$ is chosen as a null vector that satisfies $\ell^a n_a=-1$; and $m^a$, $\bar m^a$ are complex conjugate null  vector fields, taken such that their real and imaginary parts are tangential to 2-spheres (hence orthogonal to $n^a$ and $\ell^a$), and normalized as $m^a \bar m_a=1$.  In this null tetrad, the metric takes the form $\eta_{ab}=-2n_{(a} \ell_{b)}+2m_{(a} \bar m_{b)}$ and the three electromagnetic scalars are defined by:
\bea
\Phi_2 & = & F_{ab}n^a \bar m^b\, ,\\
\Phi_1 & = & \frac{1}{2} \left[ F_{ab}n^a l^b+F_{ab} m^a \bar m^b \right]  \, ,\\
\Phi_0 & = & F_{ab}m^a l^b  \, .
\eea
If we restrict to smooth solutions of Maxwell equations, the Peeling Theorem \cite{stewart94} guarantees that in a neighborhood of $\mathcal J$ we can expand $\Phi_i(u,\Omega,\theta,\phi)=\Phi_i^0 (u,0,\theta,\phi)+\Omega \, \Phi_i^1 (u,0,\theta,\phi)+ ...$, where $(u,\theta,\phi)$ are Bondi-Sachs coordinates adapted to $\mathcal J$ \cite{BMS, ae2019}. Going back to the physical spacetime, it is not difficult to see that this condition requires $\Phi_2\sim \frac{1}{r}$, so $\Phi_2^0$ represents the two radiative degrees of freedom of the electromagnetic field (corresponding to real and imaginary parts of this complex number). If we further assume the same asymptotic behaviour for the electromagnetic potential \cite{AB2017a, AB2017b}, $A_a\sim O(1/r)$, then the two radiative degrees of freedom are encoded in the component $A_2:=A_a \bar m^a$. Indeed, using $F_{ab}=2 \nabla_{[a}A_{b]}$ and the above definitions for the scalars, one can see that $A_a n^a=0$ and $\phi_2^0=\partial_u A_2^0$   at $\mathcal J$.

We are now in position to evaluate (\ref{preliminaryresult}).  The tangent space of $\mathcal J$ is spanned by $\{n^a, m^a, \bar m^a\}$, so $\epsilon^{abc}=i 3! n^{[a}m^b \bar m^{c]}$. Then,
\bea
\int_{M} p_1(F) & = &  \frac{-6i}{16\pi^2} \int_{\mathcal J} \theta_a  F_{bc}(n^{[a}m^b \bar m^{c]})  du\, d\mathbb S^2 \nonumber\\
&= & \frac{-2 i}{16\pi^2} \int_{\mathcal J} \theta_a  F_{bc}(n^{a}m^{[b} \bar m^{c]}+m^{a} \bar m^{[b} n^{c]}+\bar m^{a}n^{[b}  m^{c]} )  du\, d\mathbb S^2 \nonumber\\
&= & \frac{1}{4\pi^2} \int_{\mathcal J} du\, d\mathbb S^2\,  ( \theta_a n^a {\rm Im}\, \Phi_1^0-{\rm Im} (A_2^0 \bar\Phi_2^0 - \bar\eth \alpha\,  \bar\Phi_2^0 ) )\, ,
\eea
where we used ${A'}_2^0=\bar\eth\alpha$.  Recalling that $A_an^a=0$ at $\mathcal J$, we get
\bea
\int_{ M} p_1(F) & = &- \frac{1}{4\pi^2} \int_{\mathcal J} du\, d\mathbb S^2\,  {\rm Im} (A_2^0 \bar\Phi_2^0 - \bar\eth \alpha\,  \bar\Phi_2^0 )\, .  \label{electromagneticfinal}
 \eea
 Notice that the value of $\alpha$ is determined by the choice of gauge of $A_2^0$. In other words, under a gauge transformation $A_2^0$ transforms as $A_2 \to A_2^0+\bar \eth \beta$, while $\alpha$ transforms as $\alpha \to \alpha+\beta$, so the role of $\alpha$ is to maintain gauge invariance in the full expression. In a specific gauge, $\alpha$ can be set to zero.
 
\subsection{Physical interpretation: circularly polarized electromagnetic waves} 
 
Since $A_2^0$ (or $\Phi_2^0$) is a complex number that encodes the two radiative degrees of freedom of the electromagnetic field, we see that the chiral anomaly for fermions is intrinsically related to the emission of electromagnetic waves. Which properties should these waves have in order to produce a non-trivial result? To understand the physical meaning of this result, let us expand the electromagnetic field in Fourier modes as
\footnote{The condition $\Phi_2^0(u\to \pm \infty)\to 0$ is required from the finiteness of energy flux across $\mathcal J$, $\int_{-\infty}^{+\infty}du |\Phi(u,\theta,\phi)|^2<\infty$. This condition in turn requires that $\Phi_2^0$ belongs to $L^2(\mathbb C)$, and its Fourier transform exists. Notice that $A_2^0$ does not necessarily decay at $u\to \pm \infty$, so its Fourier transform is not defined.}
\bea
\Phi_2^0(u, \theta, \phi) & = & \int_{-\infty}^{\infty} \frac{d\omega}{2\pi} \Phi(\omega, \theta, \phi)e^{-i\omega u} = \int_0^{\infty} \frac{d\omega}{2\pi} \left[ \Phi_R(\omega, \theta, \phi)e^{-i\omega u}+ \bar \Phi_L(\omega, \theta, \phi)e^{i\omega u} \right]\, ,
\eea
where in the second equality we explicitly splitted the modes  in terms of positive and negative definite frequencies: $\Phi_R(\omega):=\Phi(\omega)$ for $\omega>0$ while $\bar \Phi_L(-\omega):=\Phi(\omega)$ for $\omega<0$. 
The electromagnetic potential  satisfies $\dot A_2=\Phi_2^0$, so we can write
\bea
A_2(u, \theta, \phi) = \int_0^{\infty} \frac{d\omega}{2\pi} \left[ \frac{\Phi_R(\omega, \theta, \phi)}{-i\omega}e^{-i\omega u}+ \frac{\bar \Phi_L(\omega, \theta, \phi)}{i\omega} e^{i\omega u} \right]+\beta(\theta,\phi)\, ,
\eea
where the function $\beta(\theta,\phi)$ emerges as a constant of integration. Imposing $\Phi_2=0$, one concludes that $\beta=\bar \eth\alpha$. 
Plugging these formulas in (\ref{electromagneticfinal}), we get
\bea
\int d^4x \sqrt{-g} \left<\nabla_a j^a_{5} \right> = \frac{e^2}{4\pi^2}\int \,d\mathbb S^2 \int_0^{\infty} \frac{d\omega}{2\pi \omega} (|\Phi_R(\omega, \theta, \phi)|^2-|\Phi_L(\omega, \theta, \phi)|^2)\, .
\eea
Note that this formula is reminiscent of the phenomenon of level-crossing, discussed in the introduction. 
What is the physical meaning of these  modes, $\Phi_R$ and $\Phi_L$? The electromagnetic field $\Phi_2^0$ is self-dual, which means that it can be written as $\Phi_2^0=( E +i B )$, where $E$, $B$ are the electric and magnetic fields, representing the two possible, linearly independent  polarization directions of the electromagnetic field. Because of this, we can also write
\bea
\Phi_2^0(u, \theta, \phi)= \int_{-\infty}^{\infty} \frac{d\omega}{2\pi}  (E(\omega, \theta, \phi) + i B(\omega, \theta, \phi))e^{-i\omega u}\, ,
\eea
from which we identify
\bea
\Phi_R(\omega) & = &   (E(\omega) + i B(\omega)), \quad \omega>0\, , \nonumber\\
\bar \Phi_L(-\omega) & = &   (E(\omega) + i B(\omega)), \quad \omega<0\, .
\eea
The second equation implies $ \Phi_L(\omega)  =    (\bar E(-\omega) - i B(-\omega))$, for $\omega>0$. Because $E(u)$, $B(u)$ are real functions, we must have then $\bar E(-\omega)=E(\omega)$ and $\bar B(-\omega)=B(\omega)$, leading to $ \Phi_L(\omega)  =    ( E(\omega) - i  B(\omega))$. Taking into account all this, and expanding the fields in spin-weight spherical harmonics of modes $(\ell,m)$, we finally arrive at
\bea
\boxed{
\left< \hat Q_5(scri+)\right> - \left< \hat Q_5(scri-)\right> = \int_0^{\infty} \frac{d\omega \, e^2}{8\pi^3 \omega} \sum_{\ell m} \left[| E^{\ell m}(\omega)+iB^{\ell m}(\omega)|^2-|E^{\ell m}(\omega) - iB^{\ell m}(\omega)|^2 \right] }\, . \nonumber
 \eea
 The RHS represents the difference in intensity between right- and left-handed circularly polarized electromagnetic waves reaching future null infinity, i.e. the Stokes V parameter. The LHS represents the amount of net helicity spontaneously created on the fermion field. We conclude that the emission of circularly polarized electromagnetic radiation  implies the spontaneous creation of massless charged fermions (highly energetic electrons, for instance) with net helicity. The more right- or left-handed electromagnetic radiation the spacetime contains, the more left or right-handed massless fermions will be excited from the quantum vacuum.

\subsection{A concrete example: electric-magnetic oscillating dipole}

Consider an   electric dipole of moment $p_0$ pointing in the z-direction,  oscillating with frequency $\omega$. The electromagnetic vector potential is \cite{Griffiths}
\bea
\vec A^E_a=-\frac{p_0 \omega}{4\pi e^2 r} \cos(\omega u) \nabla_a z\, .
\eea
On top of this, consider  a magnetic  dipole of moment $m_0$ located in the $x-y$ plane, oscillating with the same frequency $\omega$ but in opposite phase. In the radiation-zone approximation, the electromagnetic vector potential yields \cite{Griffiths} 
\bea
\vec A^M_a=-\frac{m_0 \omega}{4\pi e^2} \sin(\omega u)\sin^2\theta \nabla_a \phi\, .
\eea
One expects this configuration to provide a non trivial Chern-Pontryagin because the magnetic fields of both the electric and magnetic oscillating dipoles considered here are entangled, leading to a non-zero magnetic helicity \cite{Berger1999}. 

To calculate (\ref{electromagneticfinal}), let us work in spherical coordinates $(u,r,\theta,\phi)$. A  (conformal) Newman-Penrose basis can be constructed such that $m^a=\frac{q^{ab}}{\sqrt{2}}(\nabla_b\theta +i \sin\theta\, \nabla_a \phi)$, where $q_{ab}=\diag\{1,\sin^2\theta\}$ is the standard metric on the unit, homogeneous 2-sphere. The radiative component of the total electromagnetic potential is:
\bea
A_2^0= \bar m^a \vec A_a=\frac{\omega \sin \theta}{4\pi e^2\sqrt{2}}( i m_0\sin(\omega u)+p_0 \cos(\omega u))\, .
\eea
Then,
\bea
\int_{\mathcal J^+} du\, d\mathbb S^2\, {\rm Im} (A_2^0 \partial_u\bar A_2^0) =  -\frac{\omega^3 m_0 p_0}{16\pi e^4} (u_2-u_1)\, ,
\eea
where $ u_2-u_1$ is the period of  (finite) time during which the system operates [we let $\omega\to 0$ for $u<u_1$ and $u>u_2$]. This result is gauge-invariant, so $\alpha=0$ in (\ref{electromagneticfinal}).

\section{The gravitational case}

The fact that net fermion helicity  can be  spontaneously created from the quantum vacuum in a background of circularly polarized electromagnetic waves is  a physically interesting result. In particular, it suggests that a similar phenomenon may occur for photon helicity in a background of gravitational waves, by simply noticing the parallelism with the chiral electromagnetic anomaly of (\ref{EBanomaly}).  In this section we  prove this in detail, following a similar strategy as in the electromagnetic case. 

\subsection{Setup and main calculation}

Let ($\hat M$, $ \hat g_{ab}$)  denote our physical, curved spacetime. The quantity of interest is 
\bea
\left<\hat Q_5(scri+)\right> - \left<\hat Q_5(scri-)\right>  =\frac{\hbar}{48\pi^2}\int_{\hat M}  {\rm Tr} (\hat R\wedge \hat R )   =\frac{-\hbar}{96\pi^2} \int_{\hat M} d^4x \sqrt{-g} \hat R_{abcd}{^*\hat R}^{abcd} \, ,\label{1}
\eea
where $\hat R=\frac{1}{2}\hat R_{ab}dx^a\wedge dx^b$ is the curvature 2-form. Our goal is to compute this integral in an astrophysically relevant setting in order to know under which circumstances a given gravitational system may generate a flux of photons with net helicity. In particular, we  restrict to asymptotically flat spacetimes. As in the previous section, the above  integral is identically zero for stationary spacetimes. The proof is similar to the electromagnetic case but technically more tedious, so it is relegated to  Appendix A. We must then focus on dynamical solutions of Einstein's equations and, to guarantee convergence, we consider spacetimes that asymptotically reach stationary regimes both at future and past timelike infinities. An example of this is a binary merger of two black holes which, ideally, are initially separated an infinite distance away, and end up merging to form a final stationary Kerr black hole.  

As in the electromagnetic case, it is  convenient to work instead with a conformally compactified spacetime, ($M$, $  g_{ab}$), constructed from the physical one by the standard procedure: $M=\hat M\cup \mathcal J$, $g_{ab}=\Omega^2 \hat g_{ab}$. Our physical spacetime will be  globally hyperbolic, so that $\hat M\simeq \mathbb R \times \Sigma$ \cite{Wald84}, and in particular we shall restrict  to $\Sigma \simeq \mathbb R^3$ (physically one does not expect more sophisticated spaces). 
 The next step is to carry the relevant tensors of the physical spacetime to the unphysical one.  
It is useful to note that, due to the totally antisymmetric tensor ${\hat \epsilon}^{abmn}$, the Ricci part in the physical Riemann tensor, ${\hat R}_{abc}^{\hspace{0.5cm}d}$, does not contribute in this problem:
\bea
\sqrt{\hat g} {\hat R}_{abc}^{\hspace{0.5cm}d} {\hat \epsilon}^{abmn} {\hat R}_{mnd}^{\hspace{0.6cm}c} = \sqrt{\hat g} {\hat C}_{abc}^{\hspace{0.5cm}d} {\hat \epsilon}^{abmn} {\hat C}_{mnd}^{\hspace{0.6cm}c}  \label{fromRtoC} \, ,
\eea
where $\hat C_{abcd}$ is the physical Weyl tensor. Using now the conformal invariance of the Weyl tensor, ${\hat C}_{abc}^{\hspace{0.5cm}d}={C}_{abc}^{\hspace{0.5cm}d}$, that $\sqrt{-\hat g}=\Omega^{-4}\sqrt{-g}$, and that ${\hat \epsilon}^{abcd}=\Omega^4\epsilon^{abcd}$ 
then the quantity of interest turns out to be conformal invariant:
\bea
\sqrt{\hat g} {\hat R}_{abc}^{\hspace{0.5cm}d} {\hat \epsilon}^{abmn} {\hat R}_{mnd}^{\hspace{0.6cm}c} =\sqrt{g} {R}_{abc}^{\hspace{0.5cm}d} {\epsilon}^{abmn} {R}_{mnd}^{\hspace{0.6cm}c}  \, ,
\eea
and therefore 
\bea
\left<\hat Q_5(scri+)\right> - \left<\hat Q_5(scri-)\right>  = \frac{\hbar}{48\pi^2}\int_{\hat M}  {\rm Tr} (\hat R\wedge \hat R )   =\frac{\hbar}{48\pi^2}\int_{ M}  {\rm Tr} ( R\wedge  R )   \label{Q} \, .
\eea

Mathematically, $p_1(F)=-\frac{1}{8\pi^2} {\rm Tr} ( R\wedge  R )$ is another invariant polynomial. 
To calculate this quantity we shall recall again the Chern-Weil Theorem. This theorem tells us that the difference between two invariant polynomials, $p_1(R)-p_1(R')$, associated to any two given connection 1-forms, $\omega$ and $\omega'$ in $M$, is exact and is determined by the transgression term $Q(\omega,\omega')$ \cite{Nakahara}:
\bea
p_1(R)-p_1(R')=dQ(\omega,\omega')\, .
\eea
If we introduce the difference $\theta=\omega-\omega'$, then the RHS can be evaluated in the standard way and yields \cite{EHG}:
\bea
Q(\omega,\omega')= -\frac{1}{8\pi^2} {\rm Tr} (2\theta \wedge R+\frac{2}{3}\theta\wedge \theta\wedge \theta-2\theta \wedge\omega \wedge \theta-\theta \wedge d\theta) \, .\label{transgression}
\eea
Because the physical spacetime manifold is $\hat M\simeq \mathbb R^4$, it  admits a flat Minkowskian metric $\hat \eta_{ab}$. Its associated conformal compactification,  $(\mathbb R^4 \cup \mathcal J^+, \eta_{ab})$, with $ \eta_{ab}=\Omega^2 \hat \eta_{ab}$, will represent our auxiliary  spacetime in this calculation. We  denote   all associated quantities with prime indices. Given that ${\hat {C'}}_{abc}^{\hspace{0.5cm}d}=0$ in $(\mathbb R^4,\hat \eta_{ab})$ one has ${ C'}_{abc}^{\hspace{0.5cm}d}=0$ at any point of $\mathbb R^4 \cup \mathcal J^+$ due to conformal invariance and continuity, and  from (\ref{fromRtoC}) we deduce  $p_1(R')=p_1(C')=p_1(0)=0$. Thus, the value of $\int_{M} p_1(R)$ is simply determined by the flux $Q(\omega,\omega')$ at future null infinity (i.e. at $\Omega=0$):
\bea
\int_{M} p_1(R)=\int_{\mathcal J^+}i^*Q(\omega,\omega')= -\frac{1}{8\pi^2} \int_{\mathcal J^+} i^* {\rm Tr} (2\theta \wedge R+\frac{2}{3}\theta\wedge \theta\wedge \theta-2\theta \wedge\omega \wedge \theta-\theta \wedge d\theta) \label{firstresult0} \, .
\eea
The previous formula can be simplified in a more convenient manner. First note that $\theta \wedge d\theta=\theta \wedge d\omega - \theta \wedge d\omega'=\theta \wedge R - \theta \wedge \omega \wedge \omega +  \theta \wedge \omega' \wedge \omega'$ (recall $R'=d\omega' + \omega'\wedge \omega'=0$ for Minkowski). Then
\bea
 {\rm Tr} (-2\theta \wedge\omega \wedge \theta-\theta \wedge d\theta)  & = &  {\rm Tr} \, \theta\wedge (-2\omega\wedge\theta+\omega\wedge\omega-\omega'\wedge\omega') - {\rm Tr} \, \theta\wedge R \nonumber\\
&  = & -  {\rm Tr} \, \theta\wedge (\omega\wedge\omega-2\omega\wedge\omega'+\omega'\wedge\omega') - {\rm Tr} \, \theta\wedge R \nonumber\\
&  = & -  {\rm Tr} \, \theta\wedge ( \omega-\omega')\wedge(\omega-\omega') - {\rm Tr} \, \theta\wedge R \, ,
\eea
where in the last step we noticed that ${\rm Tr} \, \theta \wedge \omega \wedge \omega' = {\rm Tr} \, \theta \wedge \omega' \wedge \omega$.  
Eq. (\ref{firstresult0}) can now be written as
\bea
\int_{M} p_1(R)=\int_{\mathcal J^+}i^*Q(\omega,\omega')= -\frac{1}{8\pi^2} \int_{\mathcal J^+} i^*{\rm Tr} (\theta \wedge R - \frac{1}{3}\theta\wedge \theta\wedge \theta ) \, .\label{secondresult}
\eea

It is convenient to introduce  a 3+1 splitting of $(M,g_{ab})$ by $\{\Omega = const\}$ hypersurfaces in order to simplify the integrand.  Let $\hat n_a=\frac{1}{\sqrt{g^{ab}n_a n_b}}n_a$, with $n_a:=\nabla_a \Omega$, be the normalized transversal vector to the $\Omega=const$ hypersurfaces.  The induced metric on these hypersurfaces is  $h_{ab}=-\hat n_a \hat n_b+g_{ab}$, and its associated Levi-Civita derivative operator will be denoted by $D_a$. For any two vectors $u^a$, $v^a$ 
that are tangent to $\{\Omega = const\}$
we can write the decomposition:
\bea
u^aD_a v^b=u^a h^b_c \nabla_a v^c=u^a(g^b_c-\hat n^{b} \hat n_c)\nabla_a v^c=u^a\nabla_a v^b+u^a (\nabla_a \hat n_c)\hat n^{b} v^c\, , \label{extK}
\eea
where in the last equality  $v^a \hat n_a=0$ was used. This leads to $D_a v_b=h_a^c\nabla_c v_b+ \hat n_{b}v^c K_{ac}$, where $K_{ac}=D_a \hat n_c$ is the extrinsic curvature of $\{\Omega = const\}$ as a hypersurface embedded in $M$ (as usual, it satisfies $K_{ac}=K_{ca}$ and  $K_{ab}\hat n^{b}=0$, as can be easily checked).  Consider now an orthonormal frame $\{e^a_{I}\}_{I=0, \dots, 3}$  in $(M,g_{ab})$, i.e. a set of 4 vectors labelled by $I$ that at each point $x$ of $M$ satisfy $g_{ab}(x)e^a_I(x) e^b_J(x)=\eta_{IJ}$. The dual frame is defined via $e_{a,I}=g_{ab}e^b_I$, and latin indices $I, J, \dots$ can be lowered and raised with $\eta_{IJ}$. 
It is convenient to choose this frame as a Newman-Penrose tetrad ($\eta_{01}=\eta_{10}=-\eta_{23}=-\eta_{32}=-1$, zero otherwise) such that  for $\Omega=0$ the tangent space at future null infinity is spanned by $\{n^a, m^a, \bar m^a\}$.
Given this tetrad, a (torsion-free) connection 1-form $\omega_a$ is defined by the equation $\nabla_a e^I_b+\omega_{a\hspace{0.15cm}J}^{\hspace{0.15cm}I}e^J_b=0$, and the metric-compatibility condition $\nabla_a g_{bc}=0$ gives the antisymmetry property $\omega_a^{IJ}=-\omega_a^{JI}$. Taking $v_b=e_b^I$ in (\ref{extK}) we find 
\bea
h^d_a(\omega_d)_{IJ} =-e^b_J D_a e_{b I} +  K_{a I} \hat n_b e^b_J \, ,
\eea
 where $K_{a J}$ is a shorthand for  $K_{ac}e^c_J$.  Using the antisymmetry of $\omega_{a}^{IJ}$  between $I$ and $J$, one can  deduce:
 $
 h^d_a(\omega_d)^{I}_{\hspace{0.15cm}J} =-\delta^I_K(e^b_J D_a e_{b}^K) -  K_{a J} \hat n^I +  K_{a}^{I} \hat n_J \, ,
$
where $\delta_{J}^I$ and $\hat n^I$ are shorthands for $h_{ab}e^{aI} e^b_J$ and $\hat n^a e_a^I$. Let us introduce the additional  notation $(^3\omega_a)^{I}_{\hspace{0.15cm}J} \equiv -\delta^I_K(e^b_J D_a e_{b}^K)$. Thus
\bea
(\omega_a)^{I}_{\hspace{0.15cm}J} =(^3\omega_a)^{I}_{\hspace{0.15cm}J} -  K_{a J} \hat n^I+  K_{a}^{I} \hat n_J + \hat n_a \hat n^b (\omega_b)^{I}_{\hspace{0.15cm}J} \, .
\eea
 Repeating this procedure exactly in the auxiliary Minkowskian spacetime  $(\mathbb R^4 \cup \mathcal J^+, \eta_{ab})$ we get $(\omega'_a)^{I}_{\hspace{0.15cm}J} =(^3\omega'_a)^{I}_{\hspace{0.15cm}J} -  K'_{a J}\hat n'^I+  {K'}_{a}^{I} \hat n'_J +\hat n'_a \hat n'^b (\omega'_b)^{I}_{\hspace{0.15cm}J}$, where prime indices denote quantities defined with respect to the metric $\eta_{ab}$. 
 Taking the difference between the two (note that $h^a_b\hat n_a=h^a_b\hat n'_a=0$),
\bea
h^d_a(\theta_d)^{IJ} =(^3\theta_a)^{IJ} -  2(K_{a}^{[J}  \hat n^{I]}-{K'}_{a}^{[J}  \hat {n'}^{I]})  \, .
\eea

As discussed in \cite{Geroch77, Ashtekar14} and summarized in Appendix B, $\mathcal J=\{\Omega=0\}$ is a three-dimensional null hypersurface that is endowed with a universal geometric structure, which consists in a collection of pairs $(\underbar g_{ab}, \underbar n^a)$ satisfying a set of properties. Each pair consists of a degenerate metric $\underbar g_{ab}$ on $\{\Omega=0\}$ and the corresponding null normal $ \underbar n^a$. This geometric structure is common and available for any asymptotically flat spacetime. Consequently,  we can fix the same conformal frame $(\underbar g_{ab}, \underbar n^a)$  for both $(\mathbb R^4 \cup \mathcal J^+, g_{ab})$ and $(\mathbb R^4 \cup \mathcal J^+, \eta_{ab})$ spacetimes. Furthermore, without loss of generality we can choose this conformal frame such that $i^*(\Omega^{-2}  g_{ab}  n^a  n^b)=1$,  ($i^*$ being  the pullback of $i:\mathcal J\hookrightarrow M$) which will allow some simplifications in the next calculation. On the other hand, because the specification of the degenerate metric $\underbar g_{ab}$ is equivalent to the specification of two complex-conjugate vectors $m^a$, $\bar m^a$ whose real and imaginary parts are tangential to the sphere, fixing this conformal frame $(\underbar g_{ab}, \underbar n^a)$  is equivalent to fixing a common  basis $\{n^a, m^a, \bar m^a\}$ for both spacetimes. Consequently, the two tetrads introduced above, $e^a_I$ and ${e'}^a_I$, agree for $\Omega=0$. 
Taking this into account and the orthogonality properties, $(^3\omega_a)^{I}_{\hspace{0.15cm}J}  \hat n^b e_b^J=(^3\omega'_a)^{I}_{\hspace{0.15cm}J}  \hat {n'}^b {e'}_b^J= K_{a J}\hat n^b e_b^J= K'_{a J}\hat n'^b {e'}_b^J=0$, one finds
\bea
 i^*{\rm Tr}\, \theta\wedge \theta\wedge \theta & = &i^*\left[ (^3\theta_a)^{I}_{\hspace{0.15cm}J} (^3\theta_b)^{J}_{\hspace{0.15cm}K} (^3\theta_c)^{K}_{\hspace{0.15cm}I}-3 (^3\theta_a)^{IJ} (K_{bI}-{K'}_{bI})(K_{cJ}-{K'}_{cJ}) \right]\epsilon^{abc}\sqrt{h}d^3x \label{2}\, ,
\eea
and
\bea
 i^*{\rm Tr}\, \theta\wedge R & = &\frac{1}{2}  i^*\left[(^3\theta_a)^{J}_{\hspace{0.15cm}I} R_{bc\hspace{0.15cm}J}^{\hspace{0.3cm}I} -  2(K_{aI}\hat n^J -{K'}_{aI}\hat {n'}^J) R_{bc\hspace{0.15cm}J}^{\hspace{0.3cm}I}  \right] \epsilon^{abc}\sqrt{h}d^3x   \label{1} \, .
\eea

The 1-form $(^3\theta_a)^{I}_{\hspace{0.15cm}J}$ can be  determined at $\Omega=0$  from the intrinsic geometry of future null infinity. 
As discussed in Appendix B, for any covector $e_b^I$ at null infinity, the difference  $ D'_a  - D_a $ between two (equivalence classes of) connections is completely characterized by a  traceless, symmetric tensor $\sigma_{ab}$:
$
(D_a-D'_a)e_b^I =-\sigma_{ab}n^ce_c^I\, .
$
Since, as discussed above, at $\Omega=0$ the tetrad $e_a^I$  is equal to the tetrad ${e'}_a^I$, and $D_a$ is tangential to $\Omega=0$, we have $(^3\theta_a)^{I}_{\hspace{0.15cm}J}=-\delta^I_K e^b_J (D_a -D'_a)e_{b}^K$ at $\Omega=0$, and hence $(^3\theta_a)^{I}_{\hspace{0.15cm}J}e^d_I e_e^J=-e^d_I \delta^I_K  (D_a -D'_a)e_{e}^K=-e^d_I \delta^I_K \sigma_{ae} n^g e_g^K$ at $\Omega=0$. Now: $e^d_I \delta^I_K  n^g e_g^K = e^d_I (\eta^{I}_K-\hat n^I \hat n_K)  n^g e_g^K =  n^d - \hat n_g n^g \hat n^d=n^d - \frac{n^g\nabla_g \Omega g^{db}\nabla_b \Omega}{g^{ef}\nabla_e \Omega \nabla_f \Omega}$, and since at $\Omega=0$ we have $n^a=g^{ab}\nabla_b \Omega$, we conclude
\bea
i^*[(^3\theta_a)^{I}_{\hspace{0.15cm}J}e^d_I e_e^J]=0  \label{differenceD}\, ,
\eea
and we are led to
\bea
\int_{M} p_1(R)= \frac{1}{8\pi^2}  \int_{\mathcal J^+} i^* \left[ (K_{ad} -{K'}_{ad})  \hat n^e R_{bc\hspace{0.15cm}e}^{\hspace{0.25cm}d} \epsilon^{abc}\sqrt{h} d^3x  \right]  \, . \label{startingpoint}
\eea

We obtain now a compact expression for the extrinsic curvature of $\mathcal J^+$ as a hypersurface of $M$. First note that
\bea
i^*(K_{ab}-K'_{ab})=i^*(D_a \hat n_b-D'_a \hat {n'}_b)=i^*[(D_a \alpha) n_b+\alpha D_a n_b-(D'_a \alpha') n_b-\alpha' D'_a n_b]=i^*(\Omega^{-1}( D_a n_b-  D'_a n_b) ) \, ,
\eea
where we denoted $\alpha\equiv \frac{1}{\sqrt{g^{ab}\nabla_a \Omega \nabla_b \Omega}}$ for brevity and in the last step we took into account that $i^*(\Omega\alpha)=i^*(\Omega\alpha')=1$. The term inside parenthesis vanishes as $O(\Omega)$ at null infinity but the prefactor diverges as $\Omega^{-1}$, so the product is a well defined, smooth quantity at null infinity. To calculate its value let us use equation (\ref{centraleqn}) from Appendix B:
\bea
\Omega S_{ab}+2\nabla_a n_b-\Omega^{-1}n^c n_c g_{ab}=O(\Omega^{3}) \, .
\eea
The pull-back of $i: \mathcal J^+ \to M$ on this expression provides us with the value of $\Omega^{-1}\nabla_a n_b$ at future null infinity, 
\bea
i^*(\Omega^{-1}\nabla_a n_b)=-\frac{1}{2}\underbar S_{ab}+ \frac{1}{2}\underbar g_{ab} i^*(\Omega^{-2} n^an_a)=-\frac{1}{2}\underbar S_{ab}+ \frac{1}{2}\underbar g_{ab} \, .
\eea
Repeating the same with the auxiliary Minkowski space,  one gets (recall that we fixed the same conformal frame $(\underbar g_{ab}, \underbar n^c)$ for both spacetimes):
\bea
i^*(\Omega^{-1}\nabla_a n_b)-i^*(\Omega^{-1}\nabla'_a n_b) =i^*(\Omega^{-1}( D_a n_b-  D'_a n_b) )= -\frac{1}{2}\underbar (S_{ab}-S'_{ab}) \, .
\eea
But for Minkowski, $S'_{ab}=\rho_{ab}$, where $\rho_{ab}$ is the ``gauge'' part of $S_{ab}$. The combination $S_{ab}-\rho_{ab}=:N_{ab}$ is manifestly invariant under conformal gauge transformations of the form $\Omega\to \omega \Omega$,
and defines what is known as the Bondi News tensor, $N_{ab}$ (see Appendix B). This is the quantity that determines whether an asymptotically flat spacetime contains non-trivial gravitational radiation. We find
\bea
i^*(K_{ab}-K'_{ab})=-\frac{1}{2} N_{ab} \label{extrinsicK} \, .
\eea

On the other hand:
\bea
 i^*\left(\hat n^e  R_{bc\hspace{0.15cm}e}^{\hspace{0.25cm}d}\epsilon^{abc}\sqrt{h} \right)= i^*\left(n^{e}\Omega^{-1}({C}_{bce}^{\hspace{0.5cm}d}+g_{e[b}S_{c]}^d-\delta^d_{[b}S_{c]e})\epsilon^{abc}\sqrt{h} \right) \, .
\eea
Because $\hat n_a \epsilon^{abc}=0$, the second term does not contribute. On the other hand, the third term can be written as $i^*(\hat n^{e} S^n_e g_{cn}\epsilon^{adc})=i^*(g_{cn}\epsilon^{adc} \Omega^{-1})i^*( n^eS^n_e)\propto i^*(g_{cb}\epsilon^{adc} \Omega^{-1})i^*( n^b)=i^*(\epsilon^{adc})i^*(\Omega^{-1}n^bg_{bc})=0$, where we used $S^{a}_b n^b\propto n^a$ and $\underbar n^a i^*(\Omega^{-2} n_a)=1$ in the last step.
Thus,
\bea
  i^*\left(\hat n^e  R_{bc\hspace{0.15cm}e}^{\hspace{0.25cm}d}\epsilon^{abc}\sqrt{h} \right) & = & i^*\left(n^{e}\Omega^{-1} C_{bce}^{\hspace{0.5cm}d}\epsilon^{abc}\sqrt{h} \right) \nonumber\\
& = & - i^*\left( n_{e}\Omega^{-1} C_{bc}^{**\hspace{0.1cm}e d}\epsilon^{abc}\sqrt{h} \right) \nonumber\\
& = & \frac{1}{2} i^*\left( \epsilon^{dpqe}n_{e}\Omega^{-1}C^*_{bcpq} \epsilon^{abc}\sqrt{h} \, \right), \nonumber
\eea
where $*$ denotes the Hodge dual. Note that $\epsilon^{abc}=\epsilon^{abcd}\hat n_d=\Omega^{-1}\epsilon^{abcd} n_d$ but $\sqrt{h}=\Omega\sqrt{q}$, where $q_{ab}$ is the metric of the  two-dimensional spheres, so that $\epsilon^{abc}\sqrt{h}=\epsilon^{abcd} n_d\sqrt{q}$.  Now,  the quantity $(\epsilon^{dpqe}n_{e})(\epsilon^{abcm} n_m)(\Omega^{-1}C^*_{bcpq})$ is smooth in all $M$, and thus it exists in $\mathcal J^+$ (see Appendix B). Its pullback to null infinity is denoted as ${^*K}^{ad}$. Taking into account all this and using $^*K^{ab}=2\epsilon^{pq a}D_{p} N_q^{\hspace{0.15cm}b}$  \cite{Geroch77}:
\bea
i^*(K_{ad}-{K'}_{ad}) i^*\left(\hat n^e  R_{bc\hspace{0.15cm}e}^{\hspace{0.25cm}d}\epsilon^{abc}\sqrt{h} \right) & = &  - \frac{1}{4}N_{mn} {^*K}^{mn}\sqrt{q}  =  - \frac{1}{2}N_{mn} \epsilon^{pq m}D_{p} N_q^{\hspace{0.15cm}n}\sqrt{q} \, .
\eea
This expression can be simplified further. The basis  $\{n^a, m^a, \bar m^a\}$ satisfies $0=\mathcal L_n m^a=n^d D_d m^a$.
Using $\epsilon^{abc}=i 3!n^{[a}m^b \bar m^{c]}$, and $N_{ab}n^b=0$, $D_a n^b=0$, we can get
\bea
i^*(K_{ad}-{K'}_{ad}) i^*\left(\hat n^e  R_{bc\hspace{0.15cm}e}^{\hspace{0.25cm}d}\epsilon^{abc}\sqrt{h} \right) & = & -i N_{ab}  m^{[n} \bar m^{a]} n^d D_d N_n^b= - {\rm Im} (N_{44} \partial_u N_{33}) \, ,
\eea
where $N_{33}:= N_{ab}m^a m^b$ and $N_{44}=\bar N_{33}$, following the usual Newman-Penrose notation.

Taking into account these results, we can rewrite (\ref{Q}) in the final form
\bea
\int d^4x \sqrt{-g} \left<\nabla_a j^a_{5} \right> = -\frac{\hbar}{6} \int_M p_1(R)=\frac{\hbar}{48\pi^2} \int_{\mathcal J^+} du \,d\mathbb S^2  {\rm Im} (N_{44} \partial_u N_{33}) \, . \label{final2}
\eea
 Note the strong analogy with the electromagnetic case, equation (\ref{electromagneticfinal}), and also  that this result is manifestly gauge invariant ($N_{ab}$ is   invariant under conformal gauge transformations of the form $\Omega \to \omega \Omega$). 
  On the other hand, notice that this result is purely geometrical.  In other words, the topological information encoded in the Chern-Pontryagin is here trivial (zero) because we are just working with $\mathbb R^4$ with the usual differentiable structure. It is the contribution of the boundary (physically, null infinity) what contributes non-trivially to the final result, but this contribution is not topological. For manifolds with boundary, the Chern-Pontryagin is not purely topological, and its utility as a topological invariant   is recovered only when a surface correction is added \cite{EHG}. This correction is precisely equal to the result that we obtain  with a  sign reversed. 

\subsection{Physical interpretation: circularly polarized gravitational waves}

The result (\ref{final2}) tells us that the electromagnetic duality anomaly (\ref{gravcp}) is fully determined by the radiative content of the spacetime. To write the result in terms of the $\Psi_4^0$ Weyl scalar, widely used in the gravitational-wave literature,  we notice that $N_{33}=2\dot \sigma$ and $\Psi_4^0=-\ddot {\bar \sigma}$, where $\sigma$ is the shear of the gravitational radiation (see Appendix C). Then,
\bea
\int d^4x \sqrt{-g} \left<\nabla_a j^a_{5} \right> = \frac{\hbar}{12\pi^2} \int_{\mathcal J^+} du \,d\mathbb S^2 \int^{u}du' {\rm Im} ({ \Psi_4^0}(u',\theta,\phi) \bar\Psi_4^0(u,\theta,\phi))\, .  \label{final1}
\eea
Despite its apparent form, the physical interpretation of this result is remarkably simple. Expand in Fourier modes as
\bea
\Psi_4^0(u, \theta, \phi)=\int_{-\infty}^{\infty} \frac{d\omega}{2\pi}  h(\omega, \theta, \phi)e^{-i\omega u} = \int_0^{\infty} \frac{d\omega}{2\pi} \left[ h_R(\omega, \theta, \phi)e^{-i\omega u}+ \bar h_L(\omega, \theta, \phi)e^{i\omega u} \right]\, ,
\eea
where in the second equality we splitted the modes explicitly in terms of positive and negative definite modes: $h_R(\omega):=h(\omega)$ for $\omega>0$ while $\bar h_L(-\omega):=h(\omega)$ for $\omega<0$. The News scalar satisfies $\dot N_{33}=-2\bar\Psi_4^0$, so we can write
\bea
N_{33}(u, \theta, \phi) = 2\int_0^{\infty} \frac{d\omega}{2\pi} \left[ \frac{h_L(\omega, \theta, \phi)}{i\omega}e^{-i\omega u}+ \frac{\bar h_R(\omega, \theta, \phi)}{-i\omega} e^{i\omega u} \right]+\beta(\theta,\phi) \, ,
\eea
where the function $\beta(\theta,\phi)$ emerges as a constant of integration. This function, however, does not contribute to (\ref{final2}) because the physical requirement of finite  GW energy crossing null infinity implies $N_{33}(u\to\pm\infty)\to 0$ (this is inferred from the Bondi mass formula, \cite{Geroch77}), and leads to $\beta(\theta,\phi)=0$.
From (\ref{final2}),
\bea
\int d^4x \sqrt{-g} \left<\nabla_a j^a_{5} \right> = -\frac{\hbar}{12\pi^2} \int \,d\mathbb S^2 \int_0^{\infty} \frac{d\omega}{2\pi \omega} (|h_R(\omega, \theta, \phi)|^2-|h_L(\omega, \theta, \phi)|^2) \, .
\eea
What is the physical meaning of these  modes, $h_R$ and $h_L$? Because $\Psi_4^0=-\ddot {\bar \sigma} =-(\ddot h_+ - i \ddot h_{\times})$, where $h_+$, $h_{\times}$ are the two standard linear polarization modes of the GWs, we also have:
\bea
\Psi_4^0(u, \theta, \phi)= \int_{-\infty}^{\infty} \frac{d\omega}{2\pi}  \omega^2(h_+(\omega, \theta, \phi) - i h_{\times}(\omega, \theta, \phi))e^{-i\omega u} \, ,
\eea
from which we identify
\bea
h_R(\omega) & = &   \omega^2(h_+(\omega) - i h_{\times}(\omega)), \quad \omega>0\, , \nonumber\\
\bar h_L(-\omega) & = &   \omega^2(h_+(\omega) - i h_{\times}(\omega)), \quad \omega<0 \, .
\eea
The second equation implies $ h_L(\omega)  =    \omega^2(\bar h_+(-\omega) + i \bar h_{\times}(-\omega))$, for $\omega>0$. Because $h_+(u)$, $h_{\times}(u)$ are real functions, we must have $\bar h_+(-\omega)=h_+(\omega)$ and $\bar h_{\times}(-\omega)=h_{\times}(\omega)$, leading to $ h_L(\omega)  =    \omega^2( h_+(\omega) + i  h_{\times}(\omega))$, and
\bea
\boxed{
\left<Q_5(scri+)\right> - \left<Q_5(scri-)\right> = \hbar \int_0^{\infty} \frac{d\omega \omega^3}{24\pi^3} \sum_{\ell m} \left[| h_+^{\ell m}(\omega)+ih^{\ell m}_{\times}(\omega)|^2-|h_+^{\ell m}(\omega) - ih^{\ell m}_{\times}(\omega)|^2 \right] \label{FINAL}
 }\, ,
 \eea
where we expanded the field variables in spin-weight spherical harmonics of modes $(\ell,m)$. The physical interpretation of this result is again clear: while the LHS represents the net amount of photon circular polarization created,  the RHS is the difference in intensity between right- and left-handed circularly polarized GWs reaching future null infinity, i.e. the Stokes V parameter of GWs. Thus,  we conclude that the emission of chiral gravitational radiation by astrophysical systems implies the spontaneous creation of photons with net helicity. The more right- or left-handed GWs the spacetime contains, the more left or right-handed photons will be excited from the quantum vacuum.

\subsection{An example: precessing binary black hole systems}

Let us consider a binary black hole merger.  The system emits GWs, which are analyzed in modes of frequency $\omega$ and angular numbers $(\ell,m)$. During the inspiral phase the frequency spectrum is  determined by the angular velocity $\Omega$  as $\omega_m\sim m\Omega$.
The shear of the gravitational waves can be decomposed as $\sigma(u,\theta,\phi) = \sum_{ \ell m} (A^+_{\ell m}{_{-2}Y}_{\ell m}(\theta, \phi)e^{-i\omega_{m} u} + \bar A^-_{\ell m}\, {_{2}\bar Y}_{\ell m}(\theta, \phi) e^{i\omega_{m} u})$. Self-consistency requires ${\bar A}^-_{\ell m}=(-1)^m A_{\ell -m}^+$, which is deduced using ${_s\bar{Y}}_{l m}=(-1)^{s+m}{_{-s}Y}_{l(-m)}$.  Equation (\ref{FINAL}) gives
\bea
\left<Q_5(scri+)\right> - \left<Q_5(scri-)\right> & \propto & \sum_{\ell m}  \omega_m^3 ( |A^+_{\ell m}|^2- |A^+_{\ell (-m)}|^2 ) \label{exampleG} \, .
\eea
 The parameters  $A^{\pm}_{\ell m}$ can be understood as  ``excitation'' factors for the generation of each  GW polarization of mode $(\ell,m)$, and they depend on the details of the physical system under consideration (initial data).  If the binary system is invariant under mirror symmetry with respect to some plane,  choosing angular coordinates such that  $\theta=\pi/2$ represents that plane, this invariance is equivalent to say that $C_{abcd}(u,\theta,\phi)=C_{abcd}(u,\pi-\theta,\phi)$,  which implies $\sigma(u,\theta,\phi)=\bar\sigma(u,\pi-\theta,\phi)$. Using ${ }_{s} Y_{l m}(\pi-\theta, \phi+\pi)=(-1)^{l} {_{-s}Y}_{l m}(\theta, \phi)$, the previous condition leads to $A^+_{\ell m}=A^+_{\ell (-m)}(-1)^{\ell}$, which makes (\ref{exampleG}) vanish. In other words, the chiral anomaly emerges in binary mergers that do not have any mirror symmetry
\footnote{Note that  we are neglecting the backreaction of GWs on the evolution of the inspiral. When this is taken into account, the radius of the orbit shrinks for any binary merger, and  to some extent this  breaks the symmetry under spatial reversals. However,  this process can be considered adiabatic, and  its contribution insignificant.}
Examples of this are precessing binary systems, in which the individual spins of the BHs are not aligned with the total angular momentum \cite{Hannam2014}  and break any potential symmetry under mirror transformations (see \cite{dRSGMAFNS} for more details and implications in astrophysics).  

\section{Conclusions}

Chiral anomalies are a long-standing prediction of quantum field theory that have provided rich physical consequences along the last decades in several branches of physics. Despite this, their use has been considerably restricted to  non-trivial topological issues, with  instantons playing a dominant role. While this has been fruitful in many aspects, as for instance in unraveling the vacuum structure in Yang-Mills theories and solving problems of major importance in particle physics, it is not the whole story, at least in gravity (and electrodynamics). In this paper we characterized which class of solutions to Einstein's (and Maxwell) equations are able to induce the chiral anomaly on fermion and electromagnetic fields. On the one hand, we found that stationary solutions cannot trigger this anomaly. On the other hand we found that, among all dynamical solutions, only those which involve radiation with net circular polarization are able to induce the quantum anomaly, and we provided specific examples of physical interest where this occurs. 
The physical interpretation of this quantum effect  is associated to spontaneous creation of particles, but in sharp contrast to the familiar Hawking radiation of black holes, a net amount of helicity can be originated from the quantum vacuum. 
This new aspect of chiral anomalies could be useful in the search for  phenomenology, but this is out of the scope of this paper and will be left for future studies. \\

{\it Acknowledgments.--} 
 The author is grateful to I. Agullo and J. Navarro-Salas for useful comments and many discussions over the time that this work took place, and to A. Ashtekar for useful discussions on the convergence of the integrals.
 The author acknowledges support under NSF grant  PHY-1806356 and the Eberly Chair funds of Penn State; and funds from the  grant No. FIS2017-91161-EXP during an early stage of this work.

\appendix

\section{Stationary spacetimes}

In this appendix we prove that in stationary, asymptotically flat spacetimes ($M$, $g_{ab}$) with $M\simeq (t_1,t_2)\times \mathbb R^3$ one has
\bea
\int_{ M} d^4x \sqrt{-g}  R_{abcd}{^* R}^{abcd}  = 0 \, .
\eea
The argument follows in close analogy to the electromagnetic case (see section \ref{IIA}). 

Given a local orthonormal frame (``vierbein'') $\{e^{a}_I\}$, we can define the curvature 2-form from the Riemann tensor as  ${R_{ab}}^{I}_{\hspace{0.15cm}J}=R_{abc}^{\hspace{0.4cm}d}e_d^I e^c_J$. For notational simplifity we will frequently omit the internal indices $I,J$ of the curvature 2-form and/or work directly with $R=\frac{1}{2}R_{ab}\, dx^a \wedge dx^b$. If the spacetime is stationary there exists a timelike killing vector $k^a$  that leaves the metric invariant along its integral curves, $\mathcal L_{k}g_{ab}=0$.  We construct our tetrad basis $\{e^{a}_I\}$ such that $\mathcal L_k e^a_I=0$ as well.  The stationarity condition  leads to  $\mathcal L_k R_{abcd}=0$. Together with the previous equation it gives $\mathcal L_k R_{ab}=0$, or equivalently $di_k R + i_k d R=0$. For a general matrix-valued, $p$-form $V$ we can introduce the covariant derivative $DV=dV+\omega \wedge V- (-1)^pV\wedge \omega$ \cite{EHG}, under which the familiar Bianchi identity $\nabla_a {^*R}^{abcd}=0$ is equivalent to $DR=dR+\omega \wedge R-R\wedge \omega=0$. Using these equations we can write $i_k dR=i_k(-\omega\wedge R+R\wedge \omega)=-i_k\omega\wedge R+\omega \wedge i_k R+i_k R\wedge \omega+R \wedge i_k \omega$ and $d(i_k R)=D(i_kR)-\omega \wedge i_k R- i_k R \wedge \omega$. Joining both results:
\bea
D(i_k R)=i_k\omega \, R- R\,  i_k\omega \, .
\eea
For any matrix $\Lambda$ one has $D(D\Lambda)=-\Lambda R+R \Lambda$, so one can  deduce from the above that $i_k R=-D(i_k \omega)$, or $k^a R_{ab}=-\nabla_b i_k \omega$.

On the other hand, let us use the normalized vector $\hat k^a=\frac{1}{\alpha}k^a$, with $\alpha=\sqrt{-k^a k_a}$, to make a 3+1 decomposition of the metric, $g_{ab}=-\hat k_a \hat k_b+h_{ab}$. This decomposition allows the simplification: 
\bea
\int_{  M} d^4x \sqrt{-g}  R_{abcd}{^* R}^{abcd} = {\rm Tr} \int_{M} d^4x \sqrt{-g} R_{ab} {^* R}^{ab} = -4  {\rm Tr} \int_{M} d^4x \sqrt{-g} \hat k^a R_{ab} {^* R}^{\hspace{0.14cm}b}_c \hat k^c \, ,
\eea
where in the last equality we used  $^*R_{ab}=\frac{1}{2}\epsilon_{abcd}R^{cd}$ to write ${\rm Tr} \, h^{ab}h^{cd}R_{ac}{^*R_{bd}}= - 2 {\rm Tr} \, \hat k^a R_{ab} \hat k_c{^*R}^{cb}$. Doing some manipulations one gets
\bea
{\rm Tr} ((D_b i_k \omega)  \alpha^{-1}\hat k_c{^*R}^{cb}) =  \nabla_b {\rm Tr} (i_k \omega  \,\alpha^{-1} \hat k_c{^*R}^{cb}) - \nabla_b(\alpha^{-1} \hat k_c) {\rm Tr} ({^*R}^{cb} i_k \omega   ) \label{eqAux} \, .
\eea
It turns out that the second term is identically zero. To see this expand as
\bea
\nabla_b(\alpha^{-1} \hat k_c) = \alpha^{-1}\nabla_b \hat k_c-\frac{1}{\alpha^2}\hat k_c\nabla_b \alpha \, .
\eea
Since $\nabla_b \alpha=-\frac{1}{\alpha}k^d\nabla_bk_d$ and $\nabla_a k_b=- \nabla_b k_a$ ($k^a$ is a Killing Vector field):
\bea
\nabla_b(\alpha^{-1} \hat k_c) =  \alpha^{-1}( \nabla_b \hat k_c - \frac{1}{\alpha}\hat k_c\hat k^d \nabla_d k_b ) = \frac{1}{\alpha}(2  \nabla_{(b} \hat k_{c)} -D_c \hat k_b ) \, ,
\eea
where in the last equality we used $k^a\nabla_a \alpha=0$ (which can be deduced after expanding $k^a\mathcal L_k k_a=k^ag_{ab}\mathcal L_k k^b=0$), $\hat k_a \hat k_b=-g_{ab}+h_{ab}$, and introduced the spatial covariant derivative $D_a \hat k_ b=h^c_ah^d_b \nabla_c \hat k_d=h^c_a \nabla_c \hat k_b$. The RHS is a symmetric tensor, so when contracting with ${^*R}^{cb}$ in (\ref{eqAux}) the result will be zero.

We end up with the integral of a total derivative, which can be solved using Stokes Theorem. Let us work in coordinates $\{t,r, \theta, \phi\}$, where $t$ is the time   measured by static observers at spatial infinity: $k \to \frac{\partial}{\partial t}$ as $r\to \infty$. Then
\bea
\int_{M} d^4x \sqrt{-g}  R_{abcd}{^* R}^{abcd} = \lim_{r\to \infty} \int_{t_1}^{t_2} dt \int d\mathbb S^2\, r^2\, {\rm Tr} (i_k \omega   \hat k_c{^*R}^{cd}\nabla_d r)\, .
\eea
(notice that for asymptotically flat spacetimes: $\alpha\to 1$ as $r\to \infty$).  Using $\mathcal L_k e^a_I=[k,e_I]^a=0$ one can further deduce that $(i_k \omega)_{IJ}=e_I^a e_J^b\nabla_a k_b$ so
\bea
\int_{ M} d^4x \sqrt{-g}  R_{abcd}{^* R}^{abcd} =  \lim_{r\to \infty} \int_{t_1}^{t_2} dt \int d\mathbb S^2 r^2 \nabla_a k_b  {^*R}^{ ab cd}k_c\nabla_d r \, .
\eea
At spatial infinity we have $\nabla_a k_b=\nabla_a \nabla_b t=0$ so $\nabla_a k_b\sim O(r^{-1})$. Assuming standard fall-off conditions at spatial infinity for the Weyl tensor \cite{AshtekarHansen78}, $C_{abcd}\sim O(r^{-3})$, we finally see that  
\bea
\int_{M} d^4x \sqrt{-g}  R_{abcd}{^* R}^{abcd} = 0\, .
\eea

\section{Asymptotic Minkowskian spacetimes}

We summarize here the basic points of \cite{Geroch77, Ashtekar14, Ashtekar1987} that are needed to follow the  calculation in the main text.  

A spacetime $(\hat M , \hat g_{ab})$ is  called asymptotically flat at null infinity if there exists a manifold $M$ with boundary $I$ endowed with a metric tensor $g_{ab}$, and a diffeomorphism from $\hat M$ onto $M - I$ (with which we identify $\hat M$ and $M - I$)  that satisfies:\\
(a) there exists a smooth function $\Omega$ on $M$ with $g_{ab} = \Omega^2 \hat g_{ab}$ on $\hat M$; $\Omega = 0$ on $I$; and $n_a := \nabla_a \Omega$ is non-vanishing at $I$.\\
(b) $I$ is homeomorphic to $\mathbb S^2 \times \mathbb R$.\\
(c) $\tilde g_{ab}$ satisfies Einstein's equations $\hat R_{ab} - \frac{1}{2} \hat R \hat g_{ab} = 8\pi G \hat T_{ab}$, and $\Omega^{-2}\hat T_{ab}$ has
a smooth limit to $I$.

One refers to  $(\hat M , \hat g_{ab})$ as the physical spacetime, and to  $( M ,  g_{ab})$ as the unphysical one, or the conformal completion of $(\hat M , \hat g_{ab})$. Using the known conformal transformation rules for the Ricci tensor and scalar curvature, it is easy to find that these conditions imply $n^a n_a=0$ on  $I$. Thus, $I$ is a 3-dimensional null hypersurface in $M$. 

Notice that within this definition there is freedom to perform conformal rescalings: if $\Omega$ is an allowed conformal factor for a physical spacetime $(\hat M , \hat g_{ab})$, so is $\Omega' = \omega \Omega$, where $\omega$ is a smooth function on $M$ and non-vanishing at $I$. Under this {\it conformal gauge transformation}, it is easy to check that $g_{ab}\to \omega^2 g_{ab}$, $n^a \to \omega^{-1}n^a+\omega^{-2}\Omega \nabla^a \omega$. Using this freedom, it is always possible to consider a conformal completion so that $\nabla_a n^a=0$  on $I$. This gauge-fixing will be preserved under conformal gauge transformations as long as we restrict consideration to  functions $\omega$ that satisfy $n^a \nabla_a \omega=\mathcal L_n \omega=0$ on $I$. This gauge-fixing condition, together with property (c) above and the formula for the Ricci tensor under conformal transformations, implies $\nabla_a n_b=0$ on $I$, or equivalently $\nabla_a n_b=\nabla_{(a} n_{b)}=\frac{1}{2}\mathcal L_n g_{ab}=0$ on $I$. Furthermore, suppose we have any two divergence-free conformal frames associated to $\Omega$ and $\Omega'$. Because the relative conformal factor $\omega$ obeys $\mathcal L_n \omega=0$ on $I$, the vector field $n^a$ is complete if and only if $n'^a$ is complete. An asymptotically flat space-time is called asymptotically Minkowski if $I$ is complete in any divergence-free conformal frame.

Denote by $\mathcal I$ a diffeomorphic copy of $I$, and let $\xi: \mathcal I \to M$ the corresponding smooth map. The pull-back, denoted by $\xi^*$, is defined on all covariant tensor fields in $M$ in a natural way. It  can also be extended to those contravariant tensor fields such that their contraction of each of their indices with $n_a$ gives zero at $I$. Set $\underbar n^a:= \xi^*(n^a)$, $\underbar g_{ab}:=\xi^*(g_{ab})$, and $\underline \omega:=\xi^*(\omega)$. 
 It follows from the discussion above that $\mathcal I$ is endowed with the following {\it universal structure}. It is homeomorphic to $\mathbb S^2 \times \mathbb R$, and equipped with pairs of fields $(\underbar g_{ab}, \underbar n^c)$ such that: \\
 (i) $\underbar g_{ab}$ is a degenerate metric of signature $0,+,+$ with $\underbar g_{ab}\underbar n^b=0$ and $\mathcal L_{\underbar n} \underbar g_{ab}=0$; \\
 (ii) $\underbar n^a$ is complete; and, \\
 (iii) any two pairs $(\underbar g_{ab}, \underbar n^c)$ and $(\underbar g'_{ab}, \underbar n'^c)$ in the collection are related by a conformal rescaling $\Omega\to \underline\omega \Omega$ as $\underbar g'_{ab}=\underline\omega^2\underbar g_{ab}$, $\underbar n'^a=\underline\omega^{-1}\underbar n^a$, with $\mathcal L_n \underline\omega=0$.
 \footnote{The result $\xi^*(n_a)=D_a \xi^*(\Omega)= 0$ implies $\underbar g_{ab}\underbar n^b=0$. On the other hand, because the pull-back conmutes with the Lie derivative, we automatically inherit $\mathcal L_{\underbar n} \underbar g_{ab}=0$ and $\mathcal L_{\underbar n}   \underline\omega=0$.}
 
 This collection exists in any asymptotically Minwkoskian spacetime, and thereby receives the name of {\it universal structure}. A choice  of one element $(\underbar g_{ab}, \underbar n^c)$ of the collection $\{(\underbar g_{ab}, \underbar n^c)_i\}_{i\in I}$ will be called a choice of conformal frame. Note that, since 2-spheres carry a unique conformal structure, every $\underbar g_{ab}$ in this collection is conformal to a unit 2-sphere metric. Because of this, it is sometimes convenient to  restrict the remaining conformal freedom at $I$ (i.e. to fully fix the gauge function $\omega$) by demanding that the metric $q_{ab}$ on these 2-spheres be the metric of the unit radius 2-sphere. This is always possible, and this conformal frame is known as the Bondi frame.

The metric in $(M,g_{ab})$ allows the raising and lowering of indices, introduces an alternating tensor field $\epsilon^{abcd}$ unique up to a sign, and leads to a preferred derivative operator $\nabla_a$ and its associated curvature tensor $R_{abc}^{\hspace{0.4cm}d}$. Suppose we are given a fixed conformal frame. We study now what the corresponding apparatus is for $(\mathcal I, \underbar g_{ab}, \underbar n^a)$. This is not a trivial question since $\underbar g_{ab}$ is a degenerate metric. In the following we will define what fields, operations, etc one can construct  from this conformal frame, and then study their behaviour under a conformal gauge transformation. 

First of all, we can lower indices with $\underbar g_{ab}$, but we cannot raise indices a priori since $\underbar g_{ab}$ is degenerate and hence it does not have an inverse. Define a tensor $\underbar g^{ab}$ by the property: $\underbar g_{am} \underbar g^{mn}\underbar g_{nb}=\underbar g_{mn}$.This is unique up to addition of a tensor of the form $v^{(a}\underbar n^{b)}$, for any vector field $v^a$. We will use this $\underbar g^{ab}$ to raise indices whenever the lack of uniqueness does not lead to an ambiguous result. Next, we introduce an alternating tensor field $\epsilon^{abc}$, up to a sign, by the equation $\epsilon^{amn}\epsilon^{bpq}\underbar g_{mp} \underbar g_{nq}=2\underbar n^a \underbar n^b$ and demanding antisymmetry. Having fixed the sign, we can define uniquely the tensor $\epsilon_{abc}$ by $\epsilon^{abc}\epsilon_{abc}=6$ and the condition of antisymmetry. The above definition implies that $\epsilon^{abc}=\xi^*(\epsilon^{abcd}n_d)$, but note that $\epsilon_{abc}\neq \epsilon^{a'b'c'}\underbar g_{aa'}\underbar g_{bb'}\underbar g_{cc'}=\xi^*(\epsilon_{abcd}n^d)=0$. The usual identities for $\epsilon_{abc}$ and $\epsilon^{abc}$ hold.

As commented above, the universal structure of $I$ is common to {\it every} asymptotic Minkowski spacetime.  The $\mathbb S^2 \times \mathbb R$ differentiable structure together with the collection of pairs $(\underbar g_{ab}, \underbar n^c)$ is called the {\it zeroth order} structure of $I$,  and is available in any asymptotic Minkowski spacetime. We shall describe now higher  order geometrical structures that are not universal,  that contain specific physical information of the given space-time. The connection $D$ defined intrinsically on $I$ in any given conformal completion, induced by the torsion-free connection $\nabla$ compatible with $g_{ab}$, will be regarded as the first order structure. As we shall see, it contains the ``radiative information'' of the physical space-time $(\tilde M , \tilde g_{ab})$ and consequently it changes from one space-time to another.

We define the derivative operator in $\mathcal I$ by $D_a\mu_b:=\xi^*(\nabla_a \nu_b)$, where $\mu$ is any 1-form in $\mathcal I$, and $\nu_a$ is a 1-form in $M$ such that $\mu_a={\xi^*}(\nu_a)$. This derivative operator is defined intrinsically in $\mathcal I$. Notice that given any $\mu_a$ in $\mathcal I$, there exists many $\nu_a$ in $M$ that satisfies $\mu_a={\xi^*}(\nu_a)$. However, it can be shown that the derivative operator is a well-defined operation: given two $\nu_a$, $\nu'_a$ that leads to $\mu_a$ in $\mathcal I$,  one actually has $\xi^*(\nabla_a \nu_b)=\xi^*(\nabla_a \nu'_b)$. Having seen this, we can now extend the derivative operator to all tensor fields in the usual manner. In particular, given that $\nabla_a g_{bc}=0$, and $\nabla_a n^b=0$ on $I$, we find $D_{a}\underbar g_{bc}=0$, $D_a \underbar n^b=0$ (it is also not difficult to prove also that $D_a \epsilon_{bcd}=0$, $D_a \epsilon^{bcd}=0$). In other words, this derivative operator is compatible with the metric $\underbar g_{ab}$. However, it  should be remarked that this connection is  not uniquely defined because $\underbar g_{ab}$ is degenerate. We shall now characterize in physical terms the different allowed derivative operators.

 First of all, we need to know how any derivative operator changes under a conformal gauge transformation $\Omega\to \omega \Omega$. For any covector $k_a$, the transformation rule, at points of $I$, is 
 \bea
 D_{a}^{\prime} k_{b} = D_{a} k_{b}-2 \underline\omega^{-1} k_{(a} D_{b)} \underline\omega+\underline\omega^{-1}\left(\nabla^{m} \underline\omega\right) k_{m} \underbar g_{a b}\, .
 \eea
 Notice that, even when $\underline\omega=1$ so that $\underbar g_{ab}$ and $\underbar n^a$ are invariant, $D_a$ changes non-trivially as $D_{a}^{\prime} k_{b}=D_{a} k_{b}+f\left(\underbar n^{m} k_{m}\right) \underbar g_{a b}$, where we introduced $\nabla^a \underline \omega=:f \underbar n^a$. This shows that the derivative operator at $I$ is not invariant under conformal gauge transformations (in analogy to the magnetic potential in electrodynamics). Because this residual transformation of the derivative operator is just pure gauge, one is motivated to define an equivalence class of connections $\{D_a\}$, in a given conformal frame $(\underbar g_{ab}, \underbar n^c)$, by the equivalence relation:
\bea
D_a\sim D'_a \hspace{0.5cm} {\rm iff} \hspace{0.5cm}(D'_a-D_a)k_b=(f \underbar n^c k_c) \underbar g_{ab} \, , \label{equivalence}
\eea
where  $f$ is an arbitrary function on $\mathcal I$. Now, given two connections $D_a$, $D'_a$ belonging to different equivalence classes, their difference when acting on any covector is linear, and thus it must be determined by another tensor $C_{ab}^{\hspace{0.3cm}c}$: $(D_a-D'_a)k_b=C_{ab}^{\hspace{0.3cm}c} k_c$. The torsion-free derivative $\nabla_a$ implies that $C_{ab}^{\hspace{0.3cm}c}=C_{(ab)}^{\hspace{0.4cm}c}$ (just take $k_b=D_b g$ in the previous equation for some function $g$ to find $C_{ab}^{\hspace{0.3cm}c}=C_{ba}^{\hspace{0.3cm}c}$). On the other hand, the condition $D_a \underbar n^b=D'_a \underbar n^b=0$ implies $C_{ab}^{\hspace{0.3cm}c} \underbar n^b=0$, and
the metric compatibility $D_a \underbar g_{bc}=D'_a \underbar g_{bc}=0$ implies $C_{a(b}^{\hspace{0.3cm}d}\underbar g_{c)d}=0$. Since the only vector that anihilates the metric is $\underbar n^d$, then 
\bea
(D_a-D'_a)k_b=\Sigma_{ab}\underbar n^c k_c \label{differencescri} \, ,
\eea
for some tensor $\Sigma_{ab}$ with $\Sigma_{ab}=\Sigma_{(ab)}$ and $\Sigma_{ab}\underbar n^b=0$. Consequently, due to (\ref{equivalence}) the difference  $\{ D'_a \} - \{ D_a \}$ between the equivalence classes of connections is fully characterized by the trace-free tensor: $\sigma_{ab}:=\Sigma_{ab}-\frac{1}{2}\Sigma_{cd}\underbar g^{cd}\underbar g_{ab}$.  The space of equivalence classes $\{D_a\}$ is an affine space, we can select any $\{D^0_a\}$ as an origin, and then any other $\{D_a\}$ is labeled uniquely by a transverse ($\sigma_{ab}\underbar n^b=0$) trace-free symmetric tensor $\sigma_{ab}$ on $\mathcal J^+$. These properties allow to write $\sigma_{ab}=\sigma  m_a  m_b+ c.c.$, for some complex function $\sigma$. In physical terms, the two independent components of $\sigma_{ab}$ represent the two radiative degrees of freedom of the gravitational field in full general relativity.

We turn now to study the second order structure of an asymptotically Minkowski spacetime.  Let $R_{abc}^{\hspace{0.4cm}d}$ be the Riemann tensor of the unphysical spacetime, defined by $\nabla_{[a}\nabla_{b]}k_c=\frac{1}{2}R_{abc}^{\hspace{0.4cm}d} k_d$ for any covector $k_c$. The Riemann tensor can be splitted into a totally traceless part (the Weyl tensor $C_{abc}^{\hspace{0.4cm}d}$) plus a trace-full part (the Ricci $R_{ab}$, or alternatively,  Schouten tensor $S_{ab}$) as $R_{abcd}=C_{abcd}+g_{a[c}S_{d]b}-g_{b[c}S_{d]a}$. It is a fundamental result \cite{Geroch77} that the Weyl tensor vanishes at $I$, and consequently all the information about the curvature of $I$ will be determined by $S^{a}_b$. On the other hand, let us  introduce the combination $L_a^b:=\hat R_a^b-\frac{1}{6}\hat R\delta^a_b$ and $L_{ab}=g_{ac}L^{c}_b$, where $\hat R_{a}^{b}$ and $\hat R$ denote the Ricci tensor and scalar curvature of the unphysical spacetime. From the standard formula for he behaviour of the Ricci tensor under conformal transformations, and using property (c) above, one can find:\footnote{The definition of asymptotically Minkowski spacetimes requires that $\hat T_{ab}=O(\Omega^2)$, and so $\hat R_{ab}=O(\Omega^2)$.  Then $L_{ab}=g_{bc}L_a^c=O(\Omega^4)$.}
\bea
\Omega S_{ab}+2\nabla_a n_b-\Omega^{-1}n^c n_c g_{ab}=\Omega^{-1}L_{ab}=O(\Omega^3) \, . \label{centraleqn}
\eea
From this equation one deduces that, at points of $I$, $2\nabla_a n_b = 2\nabla_{(a} n_{b)}=\mathcal L_{n} g_{ab}=\Omega^{-1}n^c n_c g_{ab}$. But remember that $\mathcal L_{n} g_{ab}=0$ at points of $I$, so $f:=\Omega^{-1}n^c n_c=0$ at points of $I$ too. Now, contracting  the above equation with $n^b$ and rearranging terms, one arrives at
\bea
S_{ab}n^b+\nabla_a f=O(\Omega^2) \, .
\eea
 Since $f$ vanishes at $I$, it serves to define this hypersurface, and so its gradient must be transverse to it. Since the only trasnverse covector to $I$ is $n_a$, we necessarily have $\nabla_a f\propto n_a$. Then, $S_a^b n_b\propto n_a$, and vanishes at $I$. This means that the pull-back is well-defined on the tensor $S_a^b$, so we define  $\underbar S_a^b:=\xi^*(S_a^b)$ and also $\underbar S_{ab}: =\underbar g_{ac}\underbar S_b^c$. Notice the properties $\underbar S_{ab}\underbar n^b=0$ and $\underbar S_{ab}=\underbar S_{(ab)}$. There is one further property of $S_{ab}$ that is important  to keep in mind. By taking the pull-back of the Riemann tensor,
 and recalling the vanishing of $C_{abcd}$ at $I$, one gets
 \bea
\underbar R_{abc}^{\hspace{0.4cm}d} =\underbar g_{c[a}\underbar S_{b]}^d+\underbar S_{c[a} \underline\delta_{b]}^d \, . \label{fromRtoS}
\eea
If we define $\underbar R_{abcd}:=\underbar R_{abc}^{\hspace{0.4cm}e} \underbar g_{de}$, then the contraction of any of its indices with $\underbar n^a$ is zero.  The corresponding Ricci tensor $R_{ab}:=\underbar g^{cd} \underbar R_{abcd}$ and scalar curvature $\underbar R:=\underbar g^{cd} R_{cd}$ are thus unambiguous. Since $\underbar R_{abcd}$ lives in the 2 dimensions orthogonal to $\underbar n^a$, it can be reconstructed from its scalar curvature alone, $\underbar R_{abcd}=\underbar R \underbar g_{a[c} \underbar g_{d]b}$. Combining this with (\ref{fromRtoS}), one finally gets $\underbar g^{ab} \underbar S_{ab}=\underbar R$.

The tensor $\underbar S_{ab}$ carries information of major importance about gravitational radiation in the given spacetime, but there is still a small complication. If we change the conformal frame, this tensor transforms in a complicated way:
\bea
\underbar S_{a b}^{\prime} = \underbar S_{a b}-2 \underline\omega^{-1} D_{a} D_{b} \underline\omega+4 \underline\omega^{-2} D_{a} \underline\omega D_{b} \underline\omega-\underline\omega^{-2}\left(\underbar g^{m n} D_{m} \underline\omega D_{n} \underline\omega\right) \underbar g_{a b} \, .
\eea
Consequently, a portion of this curvature is ``gauge'' in the sense that it contains information that is not truly physical.  The goal is to extract information from this curvature tensor that remains invariant under conformal gauge transformations. This was succesfully done in \cite{Geroch77}: given any conformal frame $(\underbar g_{ab}, \underbar n^c)$, it can be proven that there exists a unique  tensor field $\rho_{ab}$ on $I$ that fulfills:
 \bea
 \rho_{[a b]}=0 , \quad \rho_{a b} \underbar n^{b}=0 , \quad \rho_{a b} \underbar g^{a b}=\underbar {R} \, , \quad D_{[a} \rho_{b] c}=0\, ,
 \eea
and, most importantly, transforms exactly as $\underbar S_{ab}$ does under a conformal gauge reescaling. Therefore, the combination 
\bea
N_{a b}:=\underbar S_{a b}-\rho_{a b}\, ,
\eea
is {\it conformally gauge invariant}. Consequently, the role of $\rho_{ab}$ is  to subtract from $S_{ab}$ the pure gauge-dependent contribution. In a Bondi conformal frame, in particular, one has $\rho_{ab}=\frac{1}{2}\underbar g_{ab}\underbar R$. $N_{ab}$ is referred to as the Bondi news tensor and is regarded as the second order structure at $\mathcal J$. It satisfies
 \bea
 N_{[a b]}=0 , \quad N_{a b} \underbar n^{b}=0 , \quad N_{a b} \underbar g^{a b}=0 \, .
  \eea
 
This is all the physical information we can extract from $S_{ab}$. Nevertheless, the full information of the curvature of $\{D_a\}$ is actually contained in  $\underbar S^a_{b}$, and not in $\underbar S_{ab}$ (notice that  since $\underbar g_{ab}$ is not invertible it is not possible to  reconstruct $\underbar S^a_{b}$ from $\underbar S_{ab}$.). This information is encoded in what we shall call the third (and last)  geometric asymptotic structure, which can be worked out from the Weyl tensor. Since the Weyl tensor vanishes at $I$, the tensor $\Omega^{-1}C_{abcd}$ is smooth up to and including $I$. If we define:
\bea
K^{ab} & := & \epsilon^{amn}\epsilon^{bpq}\xi^*(\Omega^{-1}C_{mnpq}) \, ,\\
{^*K}^{ab} & := & \epsilon^{amn}\epsilon^{bpq}\xi^*(\Omega^{-1}{^*C}_{mnpq}) \label{defK} \, ,
\eea
then we immediately see that they are symmetric and that $K^{ab}\underbar g_{ab}={^*K}^{ab}\underbar g_{ab}=0$. Taking the curl of (\ref{centraleqn}), using the definition of Riemann tensor,  expressing it in terms of the Weyl tensor, and doing some manipulations, it is possible to show that $D_{[a}\underbar S_{b]}^c=\frac{1}{4}\epsilon_{amn}{^* K}^{mc}$, which automatically leads to
\bea
D_{[a} N_{b]c}=\frac{1}{4}\epsilon_{amn}{^* K}^{mn}\underbar g_{nc} \, ,
\eea
or, equivalently,  $^*K^{ab}=2\epsilon^{pq a}D_{p} N_q^{\hspace{0.15cm}b}$. Furthermore, a straightforward calculation shows that ${^* K}^{ab}$ remains invariant under conformal gauge transformations with $\omega=1$, so it is a physically meaningful quantity.  Because ${^* K}^{ab}$ involves derivatives of $S_{a}^b$, it is called the third order structure at $I$. 

If ${^* K}^{ab}=0$, then $N_{ab}=0$, and the associated  equivalence class $\{D_a\}$ of connections is said to be trivial. In this case, the physical space-time $(\hat M , \hat g_{ab})$  does not contain gravitational radiation. In particular, every stationary, asymptotically flat  spacetime produce a trivial connection on $I$. Conversely, if $N_{ab}=0$ (i.e. no gravitational waves), it can be shown that the spacetime is stationary \cite{AshtekarXanthopoulos78}.
 
\section{Spin-coefficient  formalism and asymptotic behaviour}

Let $(M, g_{ab})$ be a spacetime and $\{\ell^a, n^a, m^a, \bar m^a\}$ a Newman-Penrose basis, i.e. a null tetrad satisfying $n^a\ell_a = 1$, $m^a \bar m_a=-1$, and zero otherwise \footnote{In this appendix and in the next one we  follow the Newman-Penrose \cite{NP62} notation. In particular, the metric signature will be $(+,-,-,-)$  in order to  use the asymptotic expressions for the spin-coefficients calculated in \cite{NU68}.} If we introduce the notation  $e_1^a=\ell^a$, $e_2^a=n^a$, $e_3^a=m^a$, $e_4^a=\bar m^a$, then this basis of null vectors  satisfies $g_{ab}=\eta_{ij}e^i_a e^j_b$ with $\eta_{12}=\eta_{21}=1$, $\eta_{34}=\eta_{43}=-1$. Internal indices ($i,j,\dots$) are raised and lowered with $\eta_{ij}$, while spacetime indices ($a,b,\dots$) are raised and lowered with $g_{ab}$. 

Given this tetrad we can introduce the connection 1-form by $\gamma_{c\hspace{0.3cm}a}^{\hspace{0.15cm}b}:=-e^i_ c\nabla_a e^b_i$, which satisfies  $\gamma_{abc}=-\gamma_{bac}$. In this basis there are 12 independent (complex) components of the connection 1-form, which are called spin coefficients. They are designated by
\bea
\kappa=\gamma_{311}=-m^a \ell^b\nabla_b\ell_a\, , \hspace{1cm} \rho=\gamma_{314}=-m^a \bar m^b\nabla_b\ell_a\, ,  \hspace{1cm} \epsilon=\frac{1}{2}( \gamma_{211}+ \gamma_{341}) =-\frac{1}{2}(n^a \ell^b \nabla_b \ell_a - \bar m^a  \ell^b \nabla_b m_a)\, ,\nonumber\\
\sigma=\gamma_{313}=-m^a m^b\nabla_b\ell_a \, , \hspace{1cm} \mu=\gamma_{243}= \bar m^a m^b\nabla_b n_a\, , \hspace{1cm} \gamma=\frac{1}{2}( \gamma_{212}+ \gamma_{342}) =-\frac{1}{2}(n^a n^b \nabla_b \ell_a - \bar m^a n^b \nabla_b m_a)\, ,\nonumber\\
\lambda=\gamma_{244}=\bar m^a \bar m^b\nabla_b n_a\, ,  \hspace{1cm} \tau=\gamma_{312}=-m^a n^b\nabla_b\ell_a\, , \hspace{1cm} \alpha=\frac{1}{2}( \gamma_{214}+ \gamma_{344}) =-\frac{1}{2}(n^a \bar m^b \nabla_b \ell_a - \bar m^a \bar m^b \nabla_b m_a)\, ,\nonumber\\
\nu=\gamma_{242}=\bar m^a n^b\nabla_b n_a \, , \hspace{1cm} \pi = \gamma_{241}=\bar m^a \ell^b\nabla_b n_a\, , \hspace{1cm} \beta=\frac{1}{2}( \gamma_{213}+ \gamma_{343})= -\frac{1}{2}(n^a m^b \nabla_b \ell_a - \bar m^a  m^b \nabla_b m_a)\, . \nonumber
\eea
Note that $\gamma_{311}=\bar\gamma_{411}$, $\gamma_{314}=\bar\gamma_{413}$, etc. On the other hand, the Weyl tensor has  ten independent components which are represented in this framework by 5 complex scalars:
\bea
\Psi_0 & := & -C_{1313}=-C_{abcd}\ell^a m^b \ell^c m^d\, , \\  
\Psi_1 & := & -C_{1213}=-C_{abcd}\ell^a n^b \ell^c m^d\, ,\\  
\Psi_2 & := & -C_{1342}=-C_{abcd}\ell^a m^b \bar m^c n^d\, ,\\  
\Psi_3 & := & -C_{1242}=-C_{abcd}\ell^a n^b \bar m^c n^d\, ,\\  
\Psi_4 & := & -C_{2424}=-C_{abcd}n^a \bar m^b n^c \bar m^d\, .
\eea
The remaining  components are determined using the symmetry properties of the Weyl tensor. In particular, it is not difficult to show that 
\bea
\Psi_1 & = & C_{1334}=C_{1231}\, ,\\
\Psi_3 & = & C_{2443}\, ,\\
{\rm Re} \Psi_2 & = & \frac{-1}{2}C_{1212}=\frac{-1}{2}C_{3434}\, ,\\
{\rm Im} \Psi_2 & = & \frac{1}{2i}C_{1234}\, ,\\
C_{1314} & = & C_{2324}=C_{1332}=C_{1442}=0\, .
\eea

In asymptotically flat spacetimes, a preferred coordinate system and an associated null tetrad can always be considered.
Following Bondi and Sachs,  we can always introduce a  foliation  of  the  asymptotic  region  of $M$ by  outgoing  null  hypersurfaces $\{u=$  const$\}$. Denoting the corresponding geodesic null normal by $\ell^a$, we can introduce an affine parameter $r$ of $\ell^a$ (i.e. $\ell=\frac{\partial}{\partial r}$ so that $\ell^a\nabla_a r=1$) such that each null surface $u=$ const is foliated by a family of (space-like) 2-spheres $\{r=$  const$\}$. The set $\{u,r,\theta,\phi\}$ is called Bondi-Sachs coordinates. Let us denote the intrinsic $(-,-)$ metric of these 2-spheres by $q_{ab}$ and the other null-normal to each of these 2-spheres by $n^a$, normalized so that $g_{ab} \ell^a n^b=1$. If $\ell^a$ is  normal to the $\{u=$  const$\}$ hypersurfaces, necessarily $\ell^a=g^{ab}\nabla_b u$, so that $\ell_a:=g_{ab}\ell^b=\nabla_a u$ and we can write  $n=V\frac{\partial}{\partial u}+U\frac{\partial}{\partial r}+X^A \frac{\partial}{\partial X^A}$ with $V=1$ ($n_a$ is not simply given by $\nabla_a r$ since $n^a$ is not normal to $\{r=$  const$\}$ hypersurfaces in general).  Finally, introduce a null complex vector field $m^a$ and its complex conjugate  $\bar m^a$ such that their real and imaginary parts are tangential to these 2-spheres,  and they are normalized such that $g_{ab} m^a \bar m^b=- 1$.  Thus,  at each point in the asymptotic region we have a null tetrad $\{\ell^a, n^a, m^a, \bar m^a\}$ for which the only non-zero contractions are $\ell\cdot n=1$ and $m \cdot \bar m=-1$. In terms of the null tetrad, the metric takes the form $g_{ab}=2n_{(a} \ell_{b)}-2m_{(a} \bar m_{b)}$. 

The spin-coefficient formalism is particularly useful for asymptotically flat spacetimes.  If  the Weyl scalars are smooth functions on the spacetime manifold, their asymptotic behaviour  as $r\to \infty$, keeping $u,\theta,\phi$ constant, is determined by the Peeling theorem  \cite{stewart94}:
\bea
\Psi_i(u,r,\theta,\phi) \sim \Psi_i^0(u,\theta,\phi)/r^{5-i}, \hspace{1cm} i=0,1,2,3,4\, .
\eea
Furthermore, the asymptotic behaviour of the spin-coefficients  can be systematically obtained by integrating asymptotically a set of equations in the Newman-Penrose framework that are equivalent to Einstein's field equations \cite{NU68}. The results read
\bea
\lambda &  = &  \lambda^0/r+O(r^{-2}),  \hspace{1cm}  \lambda^0=\dot{\bar\sigma}^0\, , \\
\mu  & =  & \mu^0 /r+O(r^{-2}),  \hspace{1cm}  \mu^0=-1\, , \\
\sigma  & =  & \sigma^0/ r^{2}+O(r^{-4}),  \hspace{1cm} \sigma^0= {\rm free \, \, data}\, , \\
\rho  & =  & \rho^0 /r+\rho^1/r^3+O(r^{-5}),  \hspace{1cm}  \rho^0=-1, \, \, \rho^1=-|\sigma^0|^2\, ,\\
\kappa &  = & 0\, ,\\
\pi & = & 0\, , \\
\nu & = & \nu^0+O(r^{-1})\, ,  \hspace{1cm}  \nu^0=0\, ,\\
\tau& = & \bar\alpha+\beta = (\bar\alpha^0+\beta^0)/r+O(r^{-2}),   \hspace{1cm} (\bar\alpha^0+\beta^0)=0\, .
\eea
and $\Psi_4^0=-\ddot{\bar\sigma}^0$. 

The relation with the Bondi News $N_{33}=N_{ab}m^a m^b$ introduced in Appendix A can be obtained using   (\ref{defK}) and the result $^*K^{ab}=2\epsilon^{pq a}D_{p} N_q^{\hspace{0.15cm}b}$. Using the first equation we get $^*K^{ab}m_a m_b=4 i n^a m^b n^c m^d \xi^*(\Omega^{-1}C_{abcd})=-4i \bar \Psi_4^0$; on the other hand, the second equation yields $^*K^{ab}m_a m_b= 2i n^p D_p (N_{ab}m^a m^b)=2i\partial_u N_{33} $. Combining both we get $\dot N_{33}=-2\bar \Psi_4^0=2\ddot \sigma^0$. Furthermore, from the Bondi mass formula $\int_{\mathcal J}du\, d\mathbb S^2 |N_{33}|^2<\infty$ one infers $N_{33}\to 0$ at $u\to \pm \infty$ so $N_{33}= 2\dot\sigma^0$.

\section{Alternative derivation using the Spin-Coefficient formalism}

In this appendix we derive the result (\ref{final2}) using the spin-coefficient formalism and the corresponding asymptotic behaviour summarized in the previous appendix. Our starting point is equation (\ref{startingpoint}), which in the physical spacetime ($ \hat M,  \hat g_{ab}$) can be rewritten in a similar way (in this appendix we only work with the physical spacetime so we will omit the hat symbol in all associated geometric quantities for convenience)
\bea
\int_{\hat M} p_1(\hat R)= -\lim_{r_0\to \infty} \frac{1}{8\pi^2}  \int_{r=r_0}  ( K_{ad} - {K'}_{ad})  \hat n^e  R_{bc\hspace{0.15cm}e}^{\hspace{0.25cm}d}  \epsilon^{abcf}\hat n_f\sqrt{ h} d^3x    \, ,
\eea
where here $ \hat n_b=\frac{1}{\sqrt{ g^{ab} \nabla_a r \nabla_b r}} \nabla_b r$ is the normal vector to $\{r=r_0\}$ hypersurfaces and $ K_{ab}=D_a  \hat n_b$ the extrinsic curvature. In terms of the Newman-Penrose basis constructed in Appendix C we must have
  $\nabla_a r=a_1n_a+a_2\ell_a$. Given that $\ell^a\nabla_a r=1$, then $\nabla_a r=n_a+a_2 \ell_a$; and  squaring $n_a=\nabla_a r-a_2\ell_a$ we get $a_2=\frac{1}{2}g^{rr}$ so that $\nabla_a r=n_a+\frac{g^{rr}}{2} \ell_a$. The asymptotic behaviour of $g^{rr}$ in Bondi-Sachs coordinates can be found in \cite{BMS, ae2019}, and is given by $-g^{rr}=1-\frac{2M}{r}+O(r^{-2})$.  Note  that $D_a  \hat n_b=h_a^{a'}h_b^{b'}\nabla_b  n_{b'}=h_a^{a'}\nabla_a \hat n_b$ and $\epsilon^{abch}\hat n_hD_a \hat n_e=\epsilon^{abch}\hat n_h\nabla_a \hat n_e$. We can also take  the prefactor of $\hat n_b$ out of the derivative operator thanks to the antisymmetry  of the Riemann tensor, 
$
(\nabla_a\hat n_e)  \hat n_d  R_{bc}^{\hspace{0.3cm}de}=\nabla_a[(g^{rr})^{-1/2} \nabla_e r ] \hat n_d  R_{bc}^{\hspace{0.3cm}de} = (g^{rr})^{-1/2} \nabla_a[ \nabla_e r ] \hat n_d  R_{bc}^{\hspace{0.3cm}de} +0
$. Finally, if we take the $\{r=r_0\}$ surface outside the gravitational sources (we assume they have spatial compact support), then $ R_{abcd}= C_{abcd}$.
Taking into account all this:
\bea
\int_{\hat M} p_1(\hat R)= \lim_{r_0\to \infty}  \frac{1}{8\pi^2} \int_{r=r_0} \frac{\sqrt{h}}{(g^{rr})^{3/2}}d^3x \,    C_{bc}^{\hspace{0.3cm}de}\epsilon^{abch} (n_h+\frac{g^{rr}}{2}l_h)(n_d+\frac{g^{rr}}{2}l_d)\nabla_a(n_e+\frac{g^{rr}}{2}l_e)\, .
\eea
We have $\nabla_a(n_e+\frac{g^{rr}}{2}l_e)=-(\gamma_{2ea}+\frac{g^{rr}}{2}\gamma_{1ea})+\frac{1}{2}l_e \nabla_a g^{rr}$. Then, using  $\epsilon^{abcd}=4! \, i\, l^{[a}n^{b}m^{c}\bar m^{d]}$, we find
\bea
\frac{-1}{8\pi^2}\int\frac{6\, \sqrt{h} d^3x}{(g^{rr})^{3/2}} \left[ ( C_{2ebc}+\frac{g^{rr}}{2} C_{1ebc})i( n^{[a}m^{b}\bar m^{c]}-\frac{g^{rr}}{2} l^{[a}m^{b}\bar m^{c]})(\gamma_{2\hspace{0.15cm}a}^{\hspace{0.15cm}e}+\frac{g^{rr}}{2}\gamma_{1\hspace{0.15cm}a}^{\hspace{0.15cm}e})- 
  C_{21bc}i( n^{[a}m^{b}\bar m^{c]}-\frac{g^{rr}}{2} l^{[a}m^{b}\bar m^{c]}) \frac{\nabla_a g^{rr}}{2} \right]
\eea
We do the calculation term by term:

(A)
\bea
 C_{2ebc} n^{[a}m^{b}\bar m^{c]} & = & \frac{1}{3} C_{2ebc} [n^{a}m^{[b}\bar m^{c]}+m^{a}\bar m^{[b} n^{c]}+\bar m^{a}n^{[b} m^{c]}]\nonumber\\
& = &\frac{1}{3} \left[n^a  C_{2e34}+m^a  C_{2e42}+\bar m^a  C_{2e23} \right] = \frac{1}{3} \left[n^a  C_{2e34} +2i {\rm Im} (m^a  C_{2e42}) \right]\, ,
 \eea

(B): same as (A) but changing $n^a\to l^a$
\bea
 C_{2ebc} l^{[a}m^{b}\bar m^{c]} & = & \frac{1}{3} C_{2ebc} [l^{a}m^{[b}\bar m^{c]}+m^{a}\bar m^{[b} l^{c]}+\bar m^{a}l^{[b} m^{c]}]\nonumber\\
& = &\frac{1}{3}\left[l^a C_{2e34}+m^a  C_{2e41}+\bar m^a  C_{2e13} \right] = \frac{1}{3}\left[l^a  C_{2e34} +2i {\rm Im} (m^a  C_{2e41}) \right]\, ,
\eea

(C): same as (A) but changing $C_{2...}\to C_{1...}$ 
\bea
 C_{1ebc} n^{[a}m^{b}\bar m^{c]} & = & \frac{1}{3} C_{1ebc} [n^{a}m^{[b}\bar m^{c]}+m^{a}\bar m^{[b} n^{c]}+\bar m^{a}n^{[b} m^{c]}]\nonumber\\
& = &\frac{1}{3}\left[n^a  C_{1e34}+m^a  C_{1e42}+\bar m^a  C_{1e23} \right] = \frac{1}{3}\left[n^a  C_{1e34}  +2i {\rm Im} (m^a  C_{1e42})\right]\, ,
\eea
 
(D): same as (B) but changing $C_{2...}\to C_{1...}$ 
\bea
 C_{1ebc} l^{[a}m^{b}\bar m^{c]} & = & \frac{1}{3} C_{1ebc} [l^{a}m^{[b}\bar m^{c]}+m^{a}\bar m^{[b} l^{c]}+\bar m^{a}l^{[b} m^{c]}]\nonumber\\
& = &\frac{1}{3}\left[l^a C_{1e34}+m^a  C_{1e41}+\bar m^a  C_{1e13} \right] = \frac{1}{3}\left[l^a  C_{1e34} +2i {\rm Im} (m^a C_{1e41})  \right]\, ,
\eea

(E): same as (A) but changing $C_{2e...}\to C_{21...}$ 
\bea
C_{21bc} n^{[a}m^{b}\bar m^{c]} = \frac{1}{3}\left[n^a C_{2134}   +2i {\rm Im} (m^a C_{2142})  \right]= \frac{1}{3}\left[-2i n^a {\rm Im} \Psi_2 +2i {\rm Im} (m^a\Psi_3)  \right]\, ,
\eea

(F): same as (B) but changing $C_{2e...}\to C_{21...}$ 
\bea
C_{21bc} l^{[a}m^{b}\bar m^{c]} =\frac{1}{3}\left[l^a C_{2134}  +2i {\rm Im} (m^a C_{2141})  \right]= \frac{1}{3}\left[-2i l^a {\rm Im} \Psi_2-2i {\rm Im} (m^a \bar\Psi_1) \right]\, .
\eea

We now elaborate in detail each of these terms:

(A)
\bea
&&\frac{1}{3}\left[n^a C_{2e34} +2i {\rm Im} (m^a C_{2e42})  \right] (\gamma_{2\hspace{0.15cm}a}^{\hspace{0.15cm}e}+\frac{g^{rr}}{2}\gamma_{1\hspace{0.15cm}a}^{\hspace{0.15cm}e})
 =  \frac{1}{3}\left[ \gamma_{2\hspace{0.15cm}2}^{\hspace{0.15cm}e}C_{2e34} 
+2i {\rm Im} (\gamma_{2\hspace{0.15cm}3}^{\hspace{0.15cm}e} C_{2e42})
   + \frac{g^{rr}}{2}\gamma_{1\hspace{0.15cm}2}^{\hspace{0.15cm}e}C_{2e34} 
   + g^{rr} i {\rm Im} ( \gamma_{1\hspace{0.15cm}3}^{\hspace{0.15cm}e} C_{2e42}) \right] \nonumber \\
&  &= \frac{1}{3}\left[  2i {\rm Im} (\gamma_{2\hspace{0.15cm}2}^{\hspace{0.15cm}3}C_{2334} ) 
+2i{\rm Im} ( \gamma_{2\hspace{0.15cm}3}^{\hspace{0.15cm}3} C_{2342} 
+ \gamma_{2\hspace{0.15cm}3}^{\hspace{0.15cm}4} C_{2442})  \right.  \nonumber\\
&& \left.+  \frac{g^{rr}}{2}(\gamma_{1\hspace{0.15cm}2}^{\hspace{0.15cm}1}C_{2134}+2 i {\rm Im} (\gamma_{1\hspace{0.15cm}2}^{\hspace{0.15cm}3}C_{2334} ) )+g^{rr} i {\rm Im}( \gamma_{1\hspace{0.15cm}3}^{\hspace{0.15cm}1} C_{2142}+  \gamma_{1\hspace{0.15cm}3}^{\hspace{0.15cm}3} C_{2342}+ \gamma_{1\hspace{0.15cm}3}^{\hspace{0.15cm}4} C_{2442})
\right] \nonumber\\
 & = & \frac{1}{3}\left[ -2i {\rm Im} (\nu \bar\Psi_3) -2i {\rm Im} (\bar \lambda \Psi_4)+2i {\rm Re}\gamma {\rm Im} \Psi_2 g^{rr}+i g^{rr} {\rm Im} (\bar\tau \bar\Psi_3) -2 i g^{rr} {\rm Re} \beta {\rm Im} ( \Psi_3)+g^{rr} i{\rm Im} (\sigma \Psi_4)\right]\, ,
\eea

(C): similar to (A) changing $C_{2...}\to C_{1...}$
\bea
&&\frac{1}{3}\left[n^a C_{1e34} +2i {\rm Im} (m^a C_{1e42}) \right] (\gamma_{2\hspace{0.15cm}a}^{\hspace{0.15cm}e}+\frac{g^{rr}}{2}\gamma_{1\hspace{0.15cm}a}^{\hspace{0.15cm}e})
 =  \frac{1}{3}\left[ \gamma_{2\hspace{0.15cm}2}^{\hspace{0.15cm}e}C_{1e34} + 2i {\rm Im} ( \gamma_{2\hspace{0.15cm}3}^{\hspace{0.15cm}e} C_{1e42})  + \frac{g^{rr}}{2}\gamma_{1\hspace{0.15cm}2}^{\hspace{0.15cm}e}C_{1e34} + g^{rr} i {\rm Im} ( \gamma_{1\hspace{0.15cm}3}^{\hspace{0.15cm}e} C_{1e42}) \right] \nonumber \\
&  &= \frac{1}{3}\left[   \gamma_{2\hspace{0.15cm}2}^{\hspace{0.15cm}2}C_{1234} +2i {\rm Im} ( \gamma_{2\hspace{0.15cm}2}^{\hspace{0.15cm}3}C_{1334} ) +2i {\rm Im} ( \gamma_{2\hspace{0.15cm}3}^{\hspace{0.15cm}2} C_{1242}+\gamma_{2\hspace{0.15cm}3}^{\hspace{0.15cm}3} C_{1342} + \gamma_{2\hspace{0.15cm}3}^{\hspace{0.15cm}4} C_{1442})  \right.  \nonumber\\
&& \left.+  \frac{g^{rr}}{2} 2 i {\rm Im} (\gamma_{1\hspace{0.15cm}2}^{\hspace{0.15cm}3}C_{1334}  )+g^{rr} i {\rm Im}(   \gamma_{1\hspace{0.15cm}3}^{\hspace{0.15cm}3} C_{1342}+ \gamma_{1\hspace{0.15cm}3}^{\hspace{0.15cm}4} C_{1442})
\right] \nonumber\\
 & = & \frac{1}{3}\left[ 4i {\rm Re \gamma} {\rm Im} \Psi_2 -2i {\rm Im} ( \nu \Psi_1)-4i {\rm Re}\beta {\rm Im} \Psi_3 +2i  {\rm Im} (\mu \Psi_2) +i g^{rr}  {\rm Im} (\bar\tau \Psi_1) - i g^{rr} {\rm Im} (\bar\rho \Psi_2)\right]\, ,
\eea

(B)
\bea
&&\frac{1}{3}\left[l^a C_{2e34} +2i {\rm Im} (m^a C_{2e41}) \right] (\gamma_{2\hspace{0.15cm}a}^{\hspace{0.15cm}e}+\frac{g^{rr}}{2}\gamma_{1\hspace{0.15cm}a}^{\hspace{0.15cm}e})
 =  \frac{1}{3}\left[ \gamma_{2\hspace{0.15cm}1}^{\hspace{0.15cm}e}C_{2e34}+2i {\rm Im} ( \gamma_{2\hspace{0.15cm}3}^{\hspace{0.15cm}e} C_{2e41})  + \frac{g^{rr}}{2}\gamma_{1\hspace{0.15cm}1}^{\hspace{0.15cm}e}C_{2e34} + g^{rr} i {\rm Im} ( \gamma_{1\hspace{0.15cm}3}^{\hspace{0.15cm}e} C_{2e41}) \right] \nonumber \\
&  &= \frac{1}{3}\left[  2i {\rm Im} (\gamma_{2\hspace{0.15cm}1}^{\hspace{0.15cm}3}C_{2334} ) + 2i {\rm Im} ( \gamma_{2\hspace{0.15cm}3}^{\hspace{0.15cm}3} C_{2341} + \gamma_{2\hspace{0.15cm}3}^{\hspace{0.15cm}4} C_{2441})  \right.  \nonumber\\
&& \left.+  \frac{g^{rr}}{2}(\gamma_{1\hspace{0.15cm}1}^{\hspace{0.15cm}1}C_{2134}+2 i {\rm Im} (\gamma_{1\hspace{0.15cm}1}^{\hspace{0.15cm}3}C_{2334} ) )+g^{rr} i {\rm Im}( \gamma_{1\hspace{0.15cm}3}^{\hspace{0.15cm}1} C_{2141}+  \gamma_{1\hspace{0.15cm}3}^{\hspace{0.15cm}3} C_{2341}+ \gamma_{1\hspace{0.15cm}3}^{\hspace{0.15cm}4} C_{2441})
\right] \nonumber\\
 & = & \frac{1}{3}\left[ -2i {\rm Im} (\pi \bar\Psi_3) +2i {\rm Im} ( \mu \bar\Psi_2)+ 2 i g^{rr} {\rm Re}\, \epsilon\,  {\rm Im} \Psi_2 + i g^{rr} {\rm im} (\bar \kappa \bar\Psi_3)+ 2 i g^{rr} {\rm Re} \beta {\rm Im} ( \bar\Psi_1)-ig^{rr}{\rm Im} (\bar\rho \bar\Psi_2)\right]\, ,
\eea

(D)
\bea
&&\frac{1}{3}\left[l^a C_{1e34} +2i {\rm Im} (m^a C_{1e41}) \right] (\gamma_{2\hspace{0.15cm}a}^{\hspace{0.15cm}e}+\frac{g^{rr}}{2}\gamma_{1\hspace{0.15cm}a}^{\hspace{0.15cm}e})
 =  \frac{1}{3}\left[ \gamma_{2\hspace{0.15cm}1}^{\hspace{0.15cm}e}C_{1e34}+2i {\rm Im} ( \gamma_{2\hspace{0.15cm}3}^{\hspace{0.15cm}e} C_{1e41})  + \frac{g^{rr}}{2}\gamma_{1\hspace{0.15cm}1}^{\hspace{0.15cm}e}C_{1e34} + g^{rr} i {\rm Im} ( \gamma_{1\hspace{0.15cm}3}^{\hspace{0.15cm}e} C_{1e41}) \right] \nonumber \\
&  &= \frac{1}{3}\left[   \gamma_{2\hspace{0.15cm}1}^{\hspace{0.15cm}2}C_{1234} + 2i {\rm Im} ( \gamma_{2\hspace{0.15cm}1}^{\hspace{0.15cm}3}C_{1334} ) + 2i {\rm Im} ( \gamma_{2\hspace{0.15cm}3}^{\hspace{0.15cm}2} C_{1241}+\gamma_{2\hspace{0.15cm}3}^{\hspace{0.15cm}3} C_{1341} + \gamma_{2\hspace{0.15cm}3}^{\hspace{0.15cm}4} C_{1441})  \right.  \nonumber\\
&& \left.+  \frac{g^{rr}}{2} 2 i {\rm Im} (\gamma_{1\hspace{0.15cm}1}^{\hspace{0.15cm}3}C_{1334}  )+g^{rr} i {\rm Im}(   \gamma_{1\hspace{0.15cm}3}^{\hspace{0.15cm}3} C_{1341}+ \gamma_{1\hspace{0.15cm}3}^{\hspace{0.15cm}4} C_{1441})
\right] \nonumber\\
 & = & \frac{1}{3}\left[ 4i {\rm Re\, \epsilon\, } {\rm Im} \Psi_2 -2i {\rm Im} ( \pi \Psi_1) + 4i {\rm Re}\beta {\rm Im} \bar\Psi_1 - 2i  {\rm Im} (\bar\lambda \bar\Psi_0) +i g^{rr}  {\rm Im} (\bar \kappa \Psi_1) + i g^{rr} {\rm Im} (\sigma \bar\Psi_0)\right]\, ,
\eea

(E)
\bea
 \frac{1}{3}\left[-2i n^a {\rm Im} \Psi_2+2i {\rm Im} (m^a \Psi_3) \right] \nabla_a g^{rr}= \frac{-2i}{3} [\frac{-2M}{r^2}+O(r^{-3})] n^a\nabla_a r {\rm Im} \Psi_2=\frac{2M i}{3 r}[1+O(r^{-1})] g^{rr} {\rm Im} \Psi_2\, ,
\eea

(F)
\bea
 \frac{1}{3}\left[-2i n^a {\rm Im} \Psi_2-2i {\rm Im} (m^a \bar\Psi_1) \right] \nabla_a g^{rr}= \frac{-2i}{3} [\frac{-2M}{r^2}+O(r^{-3})] n^a\nabla_a r {\rm Im} \Psi_2=\frac{2M i}{3 r}[1+O(r^{-1})] g^{rr} {\rm Im} \Psi_2\, .
\eea

We use now the asymptotic properties of the spin-coefficients (see Appendix C) in the limit to future null infinity to simplify all these quantities. Notice that $\sqrt{h}\sim r^2$, so all terms that decay faster than $1/r^2$ vanish at $\mathcal J^+$. On the other hand, all the spin-coefficients decay at least as $1/r$, and because all terms above are of the form ${\it spin-coefficient}$ $\times$ ${\it Weyl -scalar}$, the only non-vanishing contributions are those that involve $\Psi_4=O(\frac{1}{r})$. Among all of them, we have to take the one whose spin-coefficient only decays as $O(\frac{1}{r})$, which is $\lambda$. Doing this we get
\bea
\int d^4x \sqrt{-g} \left<\nabla_a j^a_{5} \right> = -\frac{\hbar}{6} \int_{\hat M} p_1(\hat R) =  \frac{\hbar}{12\pi^2} \int_{\mathcal J^+} du \,d\mathbb S^2 {\rm Im}\, (  \ddot \sigma^0 \dot{\bar\sigma}^0) \, .
\eea
Recalling that $ N_{33}=2\dot \sigma^0$ (see Appendix C), we recover equation (\ref{final2})

\bibliographystyle{utphys}
\bibliography{References}

\providecommand{\href}[2]{#2}\begingroup\raggedright\begin{thebibliography}{10}

\bibitem{Parker68}
L.~Parker {\em Phys. Rev. Lett.} {\bfseries 21} (1968) 562--564.

\bibitem{Hawking75}
S.~W. Hawking {\em Comm. Math. Phys.} {\bfseries 43} no.~3, (1975) 199 -- 220.

\bibitem{Bertlmann:1996xk}
R.~Bertlmann, {\em {Anomalies in quantum field theory}}.
\newblock Oxford, UK: Clarendon (1996) 566 p. (International series of
  monographs on physics: 91), 1996.

\bibitem{Adler1969}
S.~L. Adler {\em Phys. Rev.} {\bfseries 177} (1969) 2426--2438.

\bibitem{BellJackiw1969}
J.~S. Bell and R.~Jackiw {\em Il Nuovo Cimento A (1965-1970)} {\bfseries 60}
  no.~1, (1969) 47--61.

\bibitem{Nakahara}
M.~Nakahara, {\em Geometry, topology and physics}.
\newblock Bristol, UK: Hilger (1990) 505 p. (Graduate student series in
  physics).

\bibitem{Schwartz2014}
M.~D. Schwartz, {\em {Quantum Field Theory and the Standard Model}}.
\newblock Cambridge University Press, 2014.

\bibitem{Christ1980}
N.~Christ {\em Phys. Rev. D} {\bfseries 21} (1980) 1591.

\bibitem{AdRNS2017a}
I.~Agullo, A.~del Rio, and J.~Navarro-Salas {\em Phys. Rev. Lett.} {\bfseries
  118} (Mar, 2017) 111301.

\bibitem{AdRNS2017b}
I.~Agullo, A.~del Rio, and J.~Navarro-Salas {\em Int. J. Mod. Phys. D}
  {\bfseries 26} no.~12, (2017) 1742001.

\bibitem{AdRNS2018a}
I.~Agullo, A.~del Rio, and J.~Navarro-Salas {\em Phys. Rev. D} {\bfseries 98}
  no.~12, (2018) 125001.

\bibitem{AdRNS2018b}
I.~Agull\'o, A.~del R\'\i{}o, and J.~Navarro-Salas {\em Symmetry} {\bfseries
  10} no.~12, (2018) 763.

\bibitem{Kiskis1978}
J.~Kiskis {\em Phys. Rev. D} {\bfseries 18} (1978) 3690--3694.

\bibitem{Jackiw77}
R.~Jackiw {\em Rev. Mod. Phys.} {\bfseries 49} (Jul, 1977) 681--706.

\bibitem{Shifman1994}
M.~A. Shifman, ed., {\em {Instantons in gauge theories}}.
\newblock World scientific, Singapore, 1994.

\bibitem{JackiwRebbi77}
R.~Jackiw and C.~Rebbi {\em Phys. Rev. Lett.} {\bfseries 37} (1976) 172--175.

\bibitem{CallanDashenGross1976}
J.~Callan, Curtis~G., R.~Dashen, and D.~J. Gross {\em Phys. Lett. B} {\bfseries
  63} (1976) 334--340.

\bibitem{BitarChang1978}
K.~M. Bitar and S.-J. Chang {\em Phys. Rev. D} {\bfseries 17} (Jan, 1978)
  486--497.

\bibitem{tHooft76}
G.~'t~Hooft {\em Phys. Rev. Lett.} {\bfseries 37} (Jul, 1976) 8--11.

\bibitem{Hawking1977}
S.~Hawking {\em Phys. Lett. A} {\bfseries 60} no.~2, (1977) 81 -- 83.

\bibitem{EguchiHanson1979}
T.~Eguchi and A.~J. Hanson {\em Gen. Rel. and Grav.} {\bfseries 11} no.~2,
  (1979) 315 -- 320.

\bibitem{Dunajski2010}
M.~Dunajski, {\em {Solitons, instantons, and twistors}}.
\newblock 2010.

\bibitem{THOOFT1989517}
G.~{'t Hooft} {\em Nucl. Phys. B} {\bfseries 315} no.~2, (1989) 517--527.

\bibitem{Penrose1963}
R.~Penrose {\em Phys. Rev. Lett.} {\bfseries 10} (1963) 66--68.

\bibitem{Ashtekar1981}
A.~Ashtekar {\em Phys. Rev. Lett.} {\bfseries 46} (1981) 573--576.

\bibitem{Ashtekar1981b}
A.~Ashtekar {\em J. Math. Phys.} {\bfseries 22} no.~12, (1981) 2885--2895.

\bibitem{EHG}
T.~Eguchi, P.~B. Gilkey, and A.~J. Hanson, ``Gravitation, gauge theories and
  differential geometry,'' {\em Physics Reports} {\bfseries 66} no.~6, (1980)
  213 -- 393.

\bibitem{LIGOVirgoSummary}
B.~P. e.~a. Abbott {\em Phys. Rev. X} {\bfseries 9} (2019) 031040.

\bibitem{dRSGMAFNS}
A.~del Rio, N.~Sanchis-Gual, V.~Mewes, I.~Agullo, J.~A. Font, and
  J.~Navarro-Salas {\em Phys. Rev. Lett.} {\bfseries 124} (May, 2020) 211301.

\bibitem{Wald84}
R.~M. Wald, {\em {General relativity}}.
\newblock Chicago Univ. Press, Chicago, IL, 1984.

\bibitem{Geroch77}
R.~Geroch in {\em Proceedings of a Symposium on Asymptotic Structure of
  Space-Time}, F.~Esposito and L.~Witten, eds., pp.~1--105.
\newblock University of Cincinnati, Ohio, 1977.

\bibitem{Ashtekar14}
A.~Ashtekar in {\em Surveys in Differential Geometry}, L.~Bieri and S.~T.-Yau,
  eds., p.~99.
\newblock International press, Boston, 2015.

\bibitem{Ashtekar1987}
A.~Ashtekar, {\em {Asymptotic quantization: based on 1984 Naples lectures}}.
\newblock 1987.

\bibitem{NP62}
E.~Newman and R.~Penrose {\em J. Math. Phys.} {\bfseries 3} no.~3, (1962)
  566--578.

\bibitem{Chandrasekhar85}
S.~Chandrasekhar, {\em {The mathematical theory of black holes}}.
\newblock Oxford classic texts in the physical sciences. Oxford Univ. Press,
  Oxford, 2002.

\bibitem{stewart94}
J.~Stewart, {\em Advanced General Relativity}.
\newblock Cambridge Monographs on Mathematical Physics. Cambridge University
  Press, 1991.

\bibitem{BMS}
H.~Bondi, M.~G.~J. Van~der Burg, and A.~W.~K. Metzner {\em Proc. R. Soc. Lond.
  A} {\bfseries 269} no.~1336, (1962) 21--52.

\bibitem{ae2019}
F.~Alessio and G.~Esposito {\em Int. J. Geom. Methods Mod. Phys.} {\bfseries
  15} no.~02, (2018) 1830002.

\bibitem{AB2017a}
A.~Ashtekar and B.~Bonga {\em Class. Quant. Grav.} {\bfseries 34} no.~20,
  (2017) 20LT01.

\bibitem{AB2017b}
A.~Ashtekar and B.~Bonga {\em Gen. Rel. Grav.} {\bfseries 49} no.~9, (2017)
  122.

\bibitem{Griffiths}
D.~J. Griffiths, {\em {Introduction to electrodynamics; 4th ed.}}
\newblock Pearson, Boston, MA, 2013.

\bibitem{Berger1999}
M.~A. {Berger} {\em Plasma Phys. Control. Fusion} {\bfseries 41} no.~12B,
  (1999) B167--B175.

\bibitem{Hannam2014}
M.~Hannam {\em Gen. Rel. Grav.} {\bfseries 46} no.~9, (2014) .

\bibitem{AshtekarHansen78}
A.~Ashtekar and R.~O. Hansen {\em J. Math. Phys.} {\bfseries 19} (1978)
  1542--1566.

\bibitem{AshtekarXanthopoulos78}
A.~Ashtekar and B.~C. Xanthopoulos {\em J. Math. Phys.} {\bfseries 19} no.~10,
  (1978) 2216--2222.

\bibitem{NU68}
E.~T. Newman and T.~W.~J. Unti {\em J. Math. Phys.} {\bfseries 3} no.~5, (1962)
  891--901.

\end{thebibliography}\endgroup

\end{document}